\documentclass[12pt]{article}
\usepackage{bbm}
\usepackage{color}
\usepackage{authblk}
\usepackage{latexsym}
\usepackage{epsfig,amssymb,euscript, mathrsfs,cite}
\usepackage{amsmath}
\usepackage{slashed}
\usepackage{mathrsfs} 
\usepackage{amsmath}
\usepackage{dsfont}
\usepackage{mathtools}
\usepackage{enumerate}
\definecolor{MyDarkBlue}{rgb}{0.15,0.15,0.45}
\usepackage[linktocpage=true]{hyperref}
\hypersetup{
colorlinks=true,
citecolor=MyDarkBlue,
linkcolor=MyDarkBlue,
urlcolor=MyDarkBlue,
pdfauthor={},
pdftitle={},
pdfsubject={hep-th}
}
\usepackage[bbgreekl]{mathbbol}
\usepackage{bbold}
\DeclareSymbolFontAlphabet{\mathbb}{AMSb}
\DeclareSymbolFontAlphabet{\mathbbl}{bbold}
\newcommand{\spindle}{\mathbbl{\Sigma}}

\newsavebox{\ns}
\newsavebox{\dbrane}
\newsavebox{\dbshort}

\def\be{\begin{equation}}
\def\ee{\end{equation}}
\def\bea{\begin{eqnarray}}
\def\eea{\end{eqnarray}}

\newcommand{\nn}{\nonumber\\}

\newcommand\R{\mathbb{R}}
\newcommand\Z{\mathbb{Z}}

\newcommand\C{\mathbb{C}}

\newcommand\diff{\mathrm{d}}

\newcommand{\ii}{\mathrm{i}}

\newcommand{\vol}{\mathrm{vol}}

\newcommand{\x}{x}
\newcommand{\xpsi}{\psi}
\newcommand{\y}{y}
\newcommand{\yphi}{\phi}

\newlength{\sswidth}

\numberwithin{equation}{section}       

\newcommand{\uu}{u}

\begin{document}

\begin{titlepage}

\begin{flushright}
Imperial/TP/2022/JG/01\\
\end{flushright}

\vskip 1cm

\begin{center}

{\Large \bf M5-branes wrapped on four-dimensional orbifolds}

\vskip 1cm

\vskip 1cm
{K. C. Matthew Cheung$^{\mathrm{a}}$, Jacob H. T. Fry$^{\mathrm{a}}$, Jerome P. Gauntlett$^{\mathrm{a}}$,  
and James Sparks$^{\mathrm{b}}$}

\vskip 1cm

${}^{\mathrm{a}}$\textit{Blackett Laboratory, Imperial College, \\
Prince Consort Rd., London, SW7 2AZ, U.K.\\}

\vskip 0.5 cm

${}^{\,\mathrm{b}}$\textit{Mathematical Institute, University of Oxford,\\
Andrew Wiles Building, Radcliffe Observatory Quarter,\\
Woodstock Road, Oxford, OX2 6GG, U.K.\\}

\end{center}

\vskip 0.5 cm

\begin{abstract}
\noindent  
We construct supersymmetric $AdS_3$ solutions of $D=11$ supergravity, dual to $d=2$, $\mathcal{N}=(0,2)$ SCFTs,
that are associated with M5-branes wrapping two different four-dimensional orbifolds. In one case the orbifold is a
spindle fibred over another spindle, while in the other it is a spindle fibred over a Riemann surface with genus $g>1$.
We show that the central charges of the $d=2$ SCFTs calculated from the gravity solutions agree with field theory 
computations using anomaly polynomials.
The new $D=11$ solutions are obtained after constructing a new consistent Kaluza-Klein truncation of maximal $D=7$ gauged supergravity reduced on a spindle
down to $D=5$ minimal gauged supergravity.
\end{abstract}

\end{titlepage}

\pagestyle{plain}
\setcounter{page}{1}
\newcounter{bean}
\baselineskip18pt

\tableofcontents
\newpage

\section{Introduction}

The first examples of the AdS/CFT correspondence involved the near horizon limits of D3, M2 and M5-branes in flat spacetime \cite{Maldacena:1997re}. 
Following \cite{Maldacena:2000mw} it was realised that there is a rich landscape of examples that can be obtained by considering 
branes wrapping compact supersymmetric cycles in manifolds of special holonomy. In these constructions supersymmetry is preserved  
by a partial topological twist on the world-volume of the brane.

In a recent development, starting with \cite{Ferrero:2020laf}, it has been appreciated that there are more general constructions in which the branes wrap a spindle, a two-dimensional orbifold with quantised deficit angles at the two poles. The first examples
considered D3-branes wrapping spindles \cite{Ferrero:2020laf}, but constructions involving M2-branes, M5-branes and D4-branes have now also
been made \cite{Ferrero:2020twa,Hosseini:2021fge,Boido:2021szx,Ferrero:2021wvk,Couzens:2021rlk,Faedo:2021nub,Ferrero:2021etw,Giri:2021xta,Couzens:2021cpk}. These new AdS/CFT examples, which have been studied both from a gravity and a field theory point of view, have a number of interesting features. The known constructions all utilise a spindle with an azimuthal symmetry. It has recently been shown
there are then only two possibilities for preserving supersymmetry, called the ``twist" and the ``anti-twist" \cite{Ferrero:2021etw}, which are determined by the amount of magnetic R-symmetry flux threading through the spindle. While the twist is in the same topological class as the usual topological twist, it differs in the sense that the Killing spinors in the supergravity solutions are not constant on the spindle. The anti-twist case is more novel.

In the case of M5-branes and D3-branes wrapping spindles, assuming that the theory flows to a CFT in the IR, 
one can extract the central charge of the $d=4$ or $d=2$ SCFT using a-maximisation \cite{Intriligator:2003jj} or c-extremization \cite{Benini:2012cz}, respectively. One obtains the anomaly polynomial of the reduced theory by suitably integrating the anomaly polynomial of the parent theory. A novel feature
 is that the azimuthal  symmetry gives rise to a global symmetry of the reduced theory and this needs to be properly taken into account in
deriving the anomaly polynomial as discussed in \cite{Ferrero:2020laf}, extending \cite{Hosseini:2020vgl}. Another interesting aspect of branes wrapping spindles is that in some cases, involving D3-branes and M2-branes, the corresponding supergravity solutions in $D=10$ and $D=11$ are completely regular. In cases involving M5-branes, orbifold singularities remain in the $D=11$ solutions, but the precise agreement with field theory calculations strongly suggest
 that these do indeed provide {\it bona fide} examples of AdS/CFT. 
 That being said, it is an outstanding issue to determine precisely how these orbifold singularities should be treated\footnote{There has been progress in understanding such orbifold singularities in the specific context of M5-branes wrapped on a disc
 \cite{Bah:2021mzw,Bah:2021hei}.}. Finally, we also highlight that accelerating black hole solutions in $D=4$ have been given a new interpretation
 as RG flows associated with M2-branes wrapping spindles in \cite{Ferrero:2020twa} and studied further in \cite{Ferrero:2021ovq,Cassani:2021dwa}.

Given these developments, it is natural to ask if there are similar constructions involving branes wrapping higher-dimensional orbifolds.
Here we present new supersymmetric $AdS_3$ solutions of $D=11$ supergravity which describe M5-branes wrapping a particular class of four-dimensional orbifolds $M_4$ and dual to $d=2$, $\mathcal{N}=(0,2)$ SCFTs. 
The most novel construction is when $M_4=\spindle_1\ltimes\spindle_2$, which consists of
a two-dimensional spindle $\spindle_2$ that is non-trivially fibred over another spindle $\spindle_1$; a ``spindly" version of a Hirzebruch surface. We also consider another construction\footnote{This construction differs from the $AdS_3\times \spindle\times\Sigma_g$ solutions discussed in \cite{Boido:2021szx,Ferrero:2021etw} which involve a direct product of a spindle and Riemann surface (analogous solutions for D4-branes were considered in \cite{Faedo:2021nub,Giri:2021xta}).}
 when $M_4=\Sigma_g\ltimes\spindle_2$, consisting of the spindle 
$\spindle_2$ fibred over a Riemann surface $\Sigma_g$, with genus $g>1$.

We construct the new $D=11$ supergravity solutions using consistent KK truncations. Recall that there is a consistent truncation of
$D=11$ supergravity on $S^4$ down to maximal $D=7$ $SO(5)$ gauged supergravity \cite{Nastase:1999cb,Nastase:1999kf}. Here we will construct a new consistent truncation of
$D=7$ gauged supergravity on a spindle $\spindle_2$ down to minimal $D=5$ gauged supergravity. The construction is based
on the supersymmetric $AdS_5\times \spindle_2$ solution of $D=7$ gauged supergravity associated with M5-branes wrapping the
spindle $\spindle_2$ in the twist class \cite{Ferrero:2021wvk}. This $AdS_5$ solution is dual to a $d=4$, $\mathcal{N}=1$ SCFT and hence, based on the conjecture of \cite{Gauntlett:2007ma}, such a consistent truncation from $D=7$ to $D=5$ on $\spindle_2$ is to be expected.
We also show that
the $D=7$ Killing spinor equations of the $D=7$  theory reduce to those of the $D=5$ theory. This explicitly shows that any supersymmetric solution of $D=5$ minimal gauged supergravity can be uplifted on $\spindle_2$ and then on $S^4$ to
obtain supersymmetric solutions of $D=11$ supergravity.

With the consistent truncation in hand we can uplift the known $AdS_3\times \spindle_1$ solution of minimal $D=5$ gauged supergravity \cite{Ferrero:2020laf}, which is in the anti-twist class. We then obtain an $AdS_3\times \spindle_1\ltimes\spindle_2$ solution of $D=7$ gauged supergravity, which after uplifting on $S^4$ to $D=11$, describes M5-branes wrapping the four-dimensional orbifold $M_4=\spindle_1\ltimes\spindle_2$, with $\spindle_2$ non-trivially fibred over $\spindle_1$. As we will see the resulting $D=7$ solution is co-homogeneity two, but with metric functions that have a remarkable factored form which would have been
very hard to anticipate without using the consistent truncation. In a similar fashion we can also 
take the known $AdS_3\times \Sigma_g$ solution of minimal $D=5$ gauged supergravity \cite{Naka:2002jz,Klemm:2000nj}, 
which is a topological twist construction, 
and then uplift to $D=7$ to obtain $AdS_3\times \Sigma_g\ltimes\spindle_2$ solutions which are associated with 
M5-branes wrapping the four-dimensional orbifold $M_4=\Sigma_g\ltimes\spindle_2$. In both cases we calculate
the central charge from the supergravity solution and show that it precisely agrees with a field theory calculation using
anomaly polynomials and c-extremization.

\section{Consistent truncation on a spindle}
In this section we show that there is a consistent Kaluza-Klein truncation of maximal $D=7$ gauged supergravity 
on a spindle, $\spindle_2$, down to minimal $D=5$ gauged supergravity. The starting point is the $AdS_5\times \spindle_2$ solution
of \cite{Ferrero:2021wvk} which, after uplifting on an $S^4$, is dual to an $\mathcal{N}=1$ SCFT in $d=4$. 
The solution of \cite{Ferrero:2021wvk} resides in a $U(1)^2$ sub-truncation of the maximal $D=7$ gauged theory and it turns out that the consistent truncation
we are after can be formulated in this context as well.

We start with the bosonic sector of $D=7$ maximal gauged supergravity truncated to the $U(1)^2\subset SO(5)$ sector as first discussed in \cite{Liu:1999ai}. More precisely we consider $(SU(2)\times SU(2))/\mathbb{Z}_2\cong SO(4)\subset SO(5)$ and then embed the two $U(1)$'s via $[U(1)\subset SU(2)]^2$.
The bosonic field content consists of the $D=7$ metric, two gauge fields $A^{12}_{(1)}, A^{34}_{(1)}$ with field strengths
$F^{12}_{(2)}, F^{34}_{(2)}$,  two scalar fields $\lambda_i$, and a three-form $S^5_{(3)}$.
The Lagrangian is given by
\begin{align}
\mathcal{L}_{(7)}=&\,(\textit{R} -V)\text{vol}_7-6{\ast_7 d\lambda_1}\wedge d\lambda_1-6{\ast_7 d\lambda_2}\wedge d\lambda_2-8{\ast_7 d\lambda_1}\wedge d\lambda_2\nn
&-\frac{1}{2}e^{-4\lambda_1}{\ast_7 F^{12}_{(2)}}\wedge F^{12}_{(2)}-\frac{1}{2}e^{-4\lambda_2}{\ast_7 F^{34}_{(2)}}\wedge F^{34}_{(2)}-\frac{1}{2}e^{-4\lambda_1-4\lambda_2}{\ast_7 S^5_{(3)}}\wedge S^5_{(3)}\nn
&+\frac{1}{2}S^5_{(3)}\wedge dS^5_{(3)}-S^5_{(3)}\wedge F^{12}_{(2)}\wedge F^{34}_{(2)}+\frac{1}{2}A^{12}_{(1)}\wedge F^{12}_{(2)}\wedge  F^{34}_{(2)}\wedge  F^{34}_{(2)}\,,
\end{align}
where the potential is given by
\begin{align}
V=\frac{1}{2} e^{-8 (\lambda_1+\lambda_2)}-4 e^{2 (\lambda_1+\lambda_2)}-2 e^{-2 (2 \lambda_1+\lambda_2)}-2 e^{-2 (\lambda_1+2 \lambda_2)}\,.
\end{align}
We take $\epsilon_{0123456}=+1$. In appendix \ref{app:a}, we explain how this can be obtained from the maximal theory and
also present the associated Killing spinor equations\footnote{We have found many discussions in the literature 
to have typos/inconsistencies.} in~appendix~\ref{app:b}. 

\subsection{$AdS_5\times \spindle_2$ solution}\label{sec:2.1}
We first recall that this $D=7$ theory contains the supersymmetric $AdS_5\times \spindle_2$ solution found\footnote{In comparing with
\cite{Ferrero:2021wvk} we should identify $X_i=e^{2\lambda_i}$ and $A_1,A_2$ with $A^{12}_{(1)},A^{34}_{(1)}$.} in \cite{Ferrero:2021wvk}. After uplifting on $S^4$ this solution is dual to an $\mathcal{N}=1$ SCFT in $d=4$. The $D=7$ solution is given by
\begin{align}
\label{eq:M5_one_spindle_solution}
ds_7^2&=(yP)^{1/5}\left[ds^2_{AdS_5}+ds^2_{\spindle_2}\right]\,,\nn
A^{12}_{(1)}&=\frac{q_1}{h_1}d{\yphi}\,,\quad A^{34}_{(1)}=\frac{q_2}{h_2}d{\yphi}\,,\quad e^{2\lambda_i}=\frac{(yP)^{2/5}}{{h_i}}\,,
\end{align}
with vanishing three-form, $S^5_{(3)}=0$, where
$ds^2_{AdS_5}$ is a unit radius metric on $AdS_5$, 
while
\begin{align}\label{Sigmametricone}
ds^2_{\spindle_2} =   \frac{y}{4Q}dy^2+\frac{Q}{P}d\yphi ^2\, ,
\end{align}
is the metric on the spindle $\spindle_2$. The solution is specified by two real parameters
$q_1$, $q_2$ and $h_i, P$ and $Q$ are functions of $y$ given by
\begin{align}\label{eq:M5_one_spindle_solution3}
h_i(y)&=y^2+q_i\,,\nn
P(y)&=h_1(y)h_2(y)=\left(y^2+q_1\right)\left(y^2+q_2\right)\,,\nn
Q(y)&=-y^3+\frac{1}{4}P(y)=-y^3+\frac{1}{4}\left(y^2+q_1\right)\left(y^2+q_2\right)\,.
\end{align}

The Killing spinor carries charge 1/2 with respect to the gauge field $A^{12}_{(1)}+A^{34}_{(2)}$ associated with an R-symmetry of the 
$d=6$, $\mathcal{N}=(0,2)$ SCFT dual to the vacuum $AdS_7\times S^4$ solution. 
It was shown in \cite{Ferrero:2021wvk} (see also \cite{Ferrero:2021etw}) that in the
specific gauge choice in which the above $AdS_5\times \spindle_2$ solution is presented, the preserved Killing spinor has 
a phase $e^{i{\yphi}/4}$. If we perform separate  gauge transformations on the two gauge fields 
$A^{12}_{(1)}\to A^{12}_{(1)}+ c_1d{\yphi}$, $A^{34}_{(1)}\to A^{34}_{(1)}+ c_2d{\yphi}$ the phase of the Killing spinor will get modified
via $e^{i{\yphi}/4}\to  e^{i[1+2(c_1+c_2)]{\yphi}/4}$. Interestingly, 
as we will see below, in order to construct the consistent Kaluza-Klein truncation we need to utilise a specific 
gauge\footnote{Note that this is the same gauge that was used in (5.6) of \cite{Ferrero:2021etw}.} 
associated with $c_1=c_2=-1$ in which the Killing spinor has a phase $e^{-i3{\yphi}/4}$.

To ensure that $\spindle_2$ is a spindle specified by two relatively prime integers $n_\pm$ and with suitably quantised magnetic fluxes through the spindle, fixed by two integers $p_1,p_2$, it is necessary to suitably restrict the parameters $q_1,q_2$ \cite{Ferrero:2021wvk}. These solutions are necessarily in the ``twist" class \cite{Ferrero:2021etw} with
\begin{align}\label{pncons}
p_1+p_2=n_-+n_+\,.
\end{align}
 Specifically we take
\begin{align}\label{q_{1,2}}
q_1&=\frac{3\,p_1\,p_2^2\,(5n_--n_++\mathtt{s})\,(5n_+-n_-+\mathtt{s})\,(p_1-2\,p_2-\mathtt{s})\,(p_1+p_2+\mathtt{s})^2}{4\,(n_--p_1)^2\,(n_--p_2)^2\,[\mathtt{s}+2(p_1+p_2)]^4}\,, \nonumber \\
q_2&=\left. q_1\right|_{p_1 \leftrightarrow p_2}\,,
\end{align}
where 
\begin{align}\label{sdef}
\mathtt{s}\, &\equiv
\sqrt{7\,(p_1^2+p_2^2)+2\,p_1\,p_2-6\,(n_-^2+n_+^2)}\,.
\end{align}
Expressions for the 4 roots of the quartic $Q(y)$ were given in \cite{Ferrero:2021wvk} and we take 
$y$ to lie within the middle two roots
$y\in [y_2,y_3]$ where
\begin{align}\label{\y_{3}}
y_2&=\frac{3\,p_1\,p_2\,(5\,n_+-n_-+\mathtt{s})(\mathtt{s}+p_1+p_2)}{2\,(n_--p_1)(n_--p_2)[\mathtt{s}+2\,(p_1+p_2)]^2}\,, \nonumber \\
y_3&=\left. y_2\right|_{n_+\leftrightarrow n_-} \,.
\end{align}
Finally we take $\phi$ to be a periodic coordinate with period $\Delta \phi$
\begin{align}\label{Deltaz}
\frac{\Delta \yphi}{2\pi}=\frac{[\mathtt{s}-(p_1+p_2)]\,[\mathtt{s}+2(p_1+p_2)]}{9\,n_-\,n_+\,(n_--n_+)}\,.
\end{align}
This ensures that at the poles $y=y_2, y_3$ there are $\mathbb{Z}_{n_\pm}$ orbifold singularities, with conical deficit angles given by
$2\pi(1-\frac{1}{n_\pm})$, respectively.
The quantised fluxes are then given by
\begin{align}
\label{fluxcond}
\frac{1}{2\pi}\int_{\spindle_2} \diff A^{12}_{(1)} \, = \,  \frac{p_1}{n_- n_+}\,,\qquad
\frac{1}{2\pi}\int_{\spindle_2} \diff A^{34}_{(1)} \, = \,  \frac{p_2}{n_- n_+}\,,
\end{align}
with $p_i\in\mathbb{Z}$. Thus, $A^{12}_{(1)}$ and $A^{34}_{(1)}$ are connection one-forms on line bundles
$\mathcal{O}(p_1)$ and $\mathcal{O}(p_2)$ over $\spindle_2$, respectively, and, using \eqref{pncons},
the R-symmetry gauge field $A^{12}_{(1)}+A^{34}_{(1)}$ is a connection on $\mathcal{O}(n_-+n_+)$, associated with the twist
class \cite{Ferrero:2021etw} as noted above. In order to get a well defined solution one should take $n_->n_+>0$ and $p_1<0$ or $p_1>n_-+n_+$ and hence\footnote{Setting $q_1=q_2$ in the local solutions \eqref{eq:M5_one_spindle_solution}-\eqref{eq:M5_one_spindle_solution3} give
rise to local solutions in $D=7$ minimal gauged supergravity. However, the condition $p_1p_2<0$ shows that there are no spindle solutions in this sector with $p_1=p_2$.}
in particular, $p_1 p_2<0$. 

After uplifting on $S^4$, we get supersymmetric $AdS_5$ solutions to $D=11$ supergravity dual to $d=4$, $\mathcal{N}=1$ SCFTs. The corresponding central charge in the large $N$ limit calculated from the gravity solution is given by \cite{Ferrero:2021etw} 
\begin{align}\label{sugracentralcharge}
a_{4d} &  = \,\frac{3\,p_1^2\,p_2^2\,(\mathtt{s}+p_1+p_2)}{8\,n_-\,n_+\,(n_--p_1)\,(p_2-n_-)\,[\mathtt{s}+2\,(p_1+p_2)]^2}\,N^3\, ,
\end{align}
where $N$ is the quantised four-form flux though the $S^4$ and associated with the number of M5-branes wrapping the spindle.

\subsection{Consistent truncation}
We can use the solution given in \eqref{eq:M5_one_spindle_solution}-\eqref{eq:M5_one_spindle_solution3} as an inspiration to construct a consistent Kaluza-Klein truncation
on $\spindle_2$.  After some trial and error we arrived at the following ansatz.
For the $D=7$ metric we take
\begin{align}\label{const1}
ds_7^2&={(yP)^{1/5}}\Big[ds^2_{5}+\frac{y}{4Q}dy^2+\frac{Q}{P}\Big(d{\yphi}-\frac{4}{3}A_{(1)}\Big)^2\Big]\,,
\end{align}
where $ds_5^2$ is the line element for the $D=5$ metric and $A_{(1)}$ is the $D=5$ gauge field.
Furthermore, the $D=7$ gauge fields and the three-form flux are decomposed in the following way
\begin{align}\label{aessckk}
A^{12}_{(1)}&=\left(\frac{q_1}{h_1}-1\right)\left(d{\yphi}-\frac{4}{3}A_{(1)}\right)\,,\nn
A^{34}_{(1)}&=\left(\frac{q_2}{h_2}-1\right)\left(d{\yphi}-\frac{4}{3}A_{(1)}\right)\,,\nn
S^5_{(3)}&=-\frac{2y}{{3}}{\ast_5 F_{(2)}}+\frac{4yQ}{{3}h_1h_2}\left(d{\yphi}-\frac{4}{3}A_{(1)}\right)\wedge F_{(2)}\,,
\end{align} 
with the scalar fields unchanged from how they are in the $AdS_5\times \spindle_2$ solution:
\begin{align}\label{const3}
e^{2\lambda_i}=\frac{(yP)^{2/5}}{{h_i}}\,.
\end{align}

If we substitute this ansatz into the $D=7$ equations of motion we obtain
\begin{align}\label{D5EOM}
R_{\alpha\beta} & = -4g_{\alpha\beta} + \frac{2}{3}F_{\alpha\gamma}F_{\beta}^{\ \gamma} - \frac{1}{9}g_{\alpha\beta} F_{\gamma\delta}F^{\gamma\delta}\, ,\nn
d {\ast_5 F_{(2)}} & =  -\frac{2}{3}F_{(2)}\wedge F_{(2)}\, ,
\end{align}
with $\epsilon_{01234}=+1$.
These are precisely the equations of motion for $D = 5$ minimal gauged supergravity \cite{Gunaydin:1983bi} in the same
conventions as used in \cite{Ferrero:2020laf}. 

A couple of comments are in order. 
Firstly, as expected, the suitably normalised $D=5$ gauge field $A_{(1)}$ appears in the ansatz in a manner that is associated
with gauging constant shifts of the ${\yphi}$ coordinate: $d\phi\to d{\yphi}-\frac{4}{3}A_{(1)}$. 
Interestingly, we find that this needs to be done using the specific gauge choice for the gauge fields $A^{12}_{(1)}$, $A^{34}_{(1)}$ that was mentioned
just below \eqref{eq:M5_one_spindle_solution3}. 
Of course we can change the ansatz in \eqref{aessckk} by 
doing gauge transformations $A^{12}_{(1)} \to A^{12}_{(1)}+\mathfrak{a}_1d\phi$, $A^{34}_{(1)} \to A^{34}_{(1)}+\mathfrak{a}_2d\phi$
and this will lead to the same equations in \eqref{D5EOM} (as well as giving a different phase for the Killing spinor). Here we are making the point that there is a specific gauge for which $d\phi$ always appears just in the combination $d{\yphi}-\frac{4}{3}A_{(1)}$. 
One way to understand this is by noting that in the uplifted $D=11$ solution, given below in \eqref{11duplift}, if one chose a different gauge
then $\partial_\phi$ in the $D=11$ solution would not be dual\footnote{We thank the referee for this comment;
one way to directly check this would be to cast the uplifted $AdS_5\times \spindle_2$ solution in the language of \cite{Gauntlett:2004zh}.}
to the $R$-symmetry. 

Second, the consistent truncation is a local construction and is valid for any choice of the constants $q_1, q_2$. We are interested in restricting them as we discussed in the previous subsection in order that $y,\phi$ parameterise a spindle with 
suitably quantised magnetic flux. However, the consistent truncation can also be used for other values of the $q_i$ and ranges of the $y$ coordinate,
including the non-compact half spindle solutions discussed in e.g. \cite{Bah:2021mzw,Bah:2021hei},
as well as the non-compact defect solutions of \cite{Gutperle:2022pgw}.

We can also analyse the consistent truncation at the level of the Killing spinors. Specifically, in appendix \ref{app:b} we construct
an ansatz for the $D=7$ Killing spinors and show that this leads to the following Killing spinor equations for bosonic configurations of the $D=5$ theory
\begin{align}
\begin{split}\label{eq:gravitino_variation_5d_minimaltext} 
\left[{\nabla}_{\alpha}- \frac{1}{2}\beta_{\alpha}- i {}A_{\alpha}-\frac{i}{12}\left(\beta_{\alpha}^{\phantom{\alpha}\beta\rho}-4\delta_{\alpha}^\beta\beta^{\rho}\right)F_{\beta\rho}\right]\varepsilon=0\, ,
\end{split}
\end{align}
with $\beta_{01234}=-i$. This is precisely the Killing spinor equation\footnote{\label{footrchge}We highlight that in the conventions of \cite{Ferrero:2020laf} the $D=5$ supersymmetry parameters have R-charge 1 with respect to the gauge-field $A_{(1)}$. This is in contrast to charge 1/2 as in the normalisation of the gauge field used in the $D=5$ conventions of \cite{Ferrero:2021etw} and also in the $D=7$ theory, \emph{c.f}. the comment below \eqref{eq:M5_one_spindle_solution3}.} for a bosonic configuration of $D=5$ gauged supergravity satisfying
the equations of motion \eqref{D5EOM} in the conventions of \cite{Ferrero:2020laf}. This explicitly demonstrates that any supersymmetric bosonic solution
of $D=5$ minimal gauged supergravity will give rise to a supersymmetric solution of $D=7$ maximal gauged supergravity after uplifting on $\spindle_2$ via \eqref{const1}-\eqref{const3}. 
We also note that the integrability conditions for the Killing spinor
equation discussed in \cite{Gauntlett:2003fk} provide an indirect way to obtain the $D=5$ Einstein equations in \eqref{D5EOM}.

\section{$AdS_3\times  \spindle_1\ltimes \spindle_2$ solutions}
Minimal $D=5$ gauged supergravity admits a supersymmetric 
$AdS_3\times  \spindle_1$ solution, where $\spindle_1$ is a spindle \cite{Ferrero:2020laf}.
In contrast to the spindle solution discussed in the last section, which is in the twist class of \cite{Ferrero:2021etw},
these solutions are now in the anti-twist class.
Using the consistent truncation results \eqref{const1}-\eqref{const3}
we can now uplift this solution on the spindle $\spindle_2$ to obtain a supersymmetric
$AdS_3\times  \spindle_1\ltimes \spindle_2$ solution of $D=7$ gauged supergravity with $\spindle_2$ non-trivially fibred over $\spindle_1$.
This new solution is dual to a $d=2$ SCFT with $\mathcal{N}=(0,2)$ supersymmetry.

\subsection{Uplifting $D=5$ to $D=7$}

The supersymmetric $AdS_3\times \spindle_1$ solution of \cite{Ferrero:2020laf} is given by
\begin{align}\label{soln}
ds^2_5  = \frac{4\x}{9}ds^2_{AdS_3} + ds^2_{\spindle_1}\, ,\qquad
A_{(1)}  = \frac{1}{4}\left(1-\frac{a}{\x}\right)d\xpsi\,.
\end{align}
Here $ds^2_{AdS_3}$ is a unit radius metric on $AdS_3$, 
while
\begin{align}\label{Sigmametric}
ds^2_{\spindle_1} =  \frac{\x}{f}d\x^2 + \frac{f}{36 \x^2}d\xpsi^2\, ,
\end{align}
is the metric on the spindle $\spindle_1$, and $f$ is a function of $x$ given by
\begin{align}\label{efffn}
f (x)= 4\x^3 - 9\x^2 + 6a \x - a^2\, ,
\end{align}
with $a$ a constant. To ensure that $\spindle_1$ is indeed a spindle, specified by two relatively prime integers $m_\pm$, 
and with suitably quantised magnetic flux, one takes $\psi$ to be a periodic
coordinate with period $\Delta \psi$ and suitably restricts the parameter $a$:
\begin{align}
a&=\frac{(m_--m_+)^2(2m_-+m_+)^2(m_-+2 m_+)^2}{4(m_-^2+m_-m_++m_+^2)^3}\,,\nn
\Delta \psi&=\frac{2(m_-^2+m_- m_++m_+^2)}{3m_-m_+(m_-+m_+)}2 \pi\,.
\end{align}
One then takes $x\in [x_1,x_2]$ where $x_1, x_2$ are the two smallest roots of the cubic $f$. Taking $m_->m_+$ we can write
\begin{align}
x_1&=\frac{(m_--m_+)^2(m_-+2m_+)^2}{4(m_-^2+m_- m_++m_+^2)^2}\,,\nn
x_2&=\frac{(m_--m_+)^2(2m_-+m_+)^2}{4(m_-^2+m_- m_++m_+^2)^2}\, .
\end{align}
The magnetic flux through the spindle is then given by\footnote{The extra factor of 2 in the denominator as compared with \eqref{fluxcond} is due to the fact that the $D=5$ gauge field is normalised so that the supersymmetry parameters have charge 1 instead of 1/2, as noted in footnote \ref{footrchge}.}
\begin{align}\label{5dflux}
\frac{1}{2\pi}\int_{\spindle_1}F_{(2)}=\frac{m_--m_+}{2m_- m_+}\,,
\end{align}
where $F_{(2)}=dA_{(1)}$. This implies that  $2A_{(1)}$ is a connection one-form on the line bundle $\mathcal{O}(m_--m_+)$ over the spindle $\spindle_1$ and hence we are in the anti-twist class as noted above.

Using the ansatz given in \eqref{const1}-\eqref{const3} we can now obtain the $AdS_3\times  \spindle_1\ltimes \spindle_2$ solution of $D=7$ gauged
supergravity. The $D=7$ metric is given by
\begin{align}\label{d7sssol}
ds_7^2
&={(yP)^{1/5}} \frac{4{\x}}{9}\Big[ds^2_{AdS_3} + \frac{9}{4{f}}d{\x}^2 + \frac{{f}}{16{\x}^3}d{\xpsi}^2+\frac{9}{16{\x}}\frac{y}{Q}dy^2 \nonumber\\
& \qquad \qquad \qquad +\frac{9}{4{\x}}\frac{Q}{P}\Big(d\yphi-\frac{1}{3}\left(1-\frac{a}{{\x}}\right)d{\xpsi}\Big)^2\Big]\,,
\end{align}
while the remaining fields are given by
\begin{align}\label{d7sssol2}
A^{12}_{(1)}&=\left(\frac{q_1}{h_1}-1\right)\left(d\yphi-\frac{1}{3}\left(1-\frac{a}{{\x}}\right)d{\xpsi}\right)\,,\nn
A^{34}_{(1)}&=\left(\frac{q_2}{h_2}-1\right)\left(d\yphi-\frac{1}{3}\left(1-\frac{a}{{\x}}\right)d{\xpsi}\right)\,,\nn
S^5_{(3)}&=-\frac{8ay}{27}\text{vol}(AdS_3)+\frac{ay Q}{3 x^2 h_1 h_2}d{\x}\wedge  d{\xpsi}\wedge d\yphi\,,\nn
e^{2\lambda_i}&=\frac{(yP)^{2/5}}{{h_i}}\,.
\end{align}
Recall that $f=f(x)$ while $Q,P,h_i$ are all functions of $y$. The four-dimensional internal space metric in \eqref{d7sssol} has two Killing vectors $\partial_\phi$ and $\partial_\psi$. It is thus co-homogeneity two but we observe
that there is a remarkable separation of variables in the metric functions, including the overall warp factor. 

Clearly the internal space has the form of the spindle $\spindle_2$, parametrised by $(y,\phi)$, fibred over the spindle $\spindle_1$, parametrised by $(x,\psi)$. To ensure that this fibration is well defined (in the orbifold sense), we demand that the
one-form determining the fibration, $\eta\equiv \frac{2\pi}{\Delta\phi}(d\phi-\frac{1}{3}\left(1-\frac{a}{{\x}}\right)d{\xpsi})$, is globally defined. This
requires that
\begin{align}\label{tqncond1}
\frac{1}{2\pi} \int_{\spindle_1}d\eta= \frac{{t}}{m_-m_+}\, ,\qquad {t}\in \mathbb{Z}\,.
\end{align}
But since $d\eta= -\frac{2\pi}{\Delta \yphi}\frac{4}{3} F_{(2)}$, using \eqref{Deltaz} and \eqref{5dflux} we immediately deduce that
we need to impose the following condition on the two sets of spindle quantum numbers $m_\pm, n_\pm$ as well as the $p_i$ satisfying
\eqref{pncons}:
\begin{align}\label{tcond1}
 {t}=-{6}(m_--m_+)\frac{\,n_-\,n_+\,(n_--n_+)}{[\mathtt{s}-(p_1+p_2)]\,[\mathtt{s}+2(p_1+p_2)]}\in \mathbb{Z}\,.
\end{align}
This condition ensures that away from the poles on the $\spindle_2$ fibre the space $\spindle_1\ltimes\spindle_2$
is well-defined, and moreover when $t$ is relatively prime to both $m_\pm$ this total space is smooth, including at the poles of the $\spindle_1$ spindle base. Indeed at constant value of $y\ne y_2,y_3$, the total space parametrised by the $\spindle_1$ base and the circle parametrised by $\phi$ will
then be a Lens space (see appendix A of \cite{Ferrero:2020twa}). 
 However, there are orbifold singularities associated with the two poles of the $\spindle_2$ fibre, when
$y=y_2,y_3$. The resulting four-dimensional space $M_4$ is then a spindly version of a Hirzebruch surface. 
We may describe this more globally by starting with the base spindle $\spindle_1=\mathbb{WCP}^1_{[m_-,m_+]}$, 
together with the $U(1)$ orbibundle $\mathcal{O}(t)$ over it (with $e^{2\pi \ii \phi/\Delta\phi}$ the fibre coordinate), where by definition the first Chern class is given 
by  \eqref{tqncond1}.  One then uses the transition functions for this bundle to fibre 
$\spindle_2=\mathbb{WCP}^1_{[n_-,n_+]}$ over $\spindle_1$, with $U(1)$ acting on the fibres $\spindle_2$ 
by rotation, fixing the poles.\footnote{Here one should also be careful to use an appropriate local model for the fibration 
near to the two poles of the base $\spindle_1$, as described in detail in \cite{Ferrero:2021etw}. Specifically, 
near such a pole of $\spindle_1$,  $M_4$ is modelled as a $\Z_{m_\pm}$ quotient of $\C\times \spindle_2$, where $\Z_{m_\pm}$ acts 
on both factors in this product.}
Notice here that the twisting parameter $t\in\Z$ can in principle be arbitrary, but 
that for the particular solutions we have constructed this is fixed in terms of other parameters via 
\eqref{tcond1}. This is likely to be an artefact of the particular ansatz we have taken for constructing the $D=11$ solutions
using a double uplift/consistent truncation; going outside this framework should allow for solutions with more general values for
the parameter $t$ . We also note that the resulting space $M_4$ is naturally a toric complex orbifold, 
and as such can also be described by a gauged linear sigma model (GLSM). Specifically, 
$M_4$ may be realised as the vacuum moduli space of a $U(1)^2$ theory with 
4 complex fields of charges $(0,-t,m_-,m_+)$ and $(n_+,n_-,0,0)$.

We will not attempt to find the general solution to \eqref{tcond1} here since, as we will see in the next section, additional conditions also
need to be imposed for regularity when uplifting to $D=11$. Nevertheless, we can use the results of
\cite{Ferrero:2021wvk} to show that solutions do exist. Specifically, we recall the generating formula given in \cite{Ferrero:2021wvk},
which gives an infinite subset of $AdS_5\times \spindle_2$ solutions:
\begin{align}\label{pkfamily}
p_1 = \frac{n_-+n_+}{2}-\frac{3n_--n_+}{4}(\beta_+^k + \beta_-^k) - \frac{5n_--n_+}{4\sqrt{3}}(\beta_+^k-\beta_-^k)\, .
\end{align}
Here $k\in\Z_{\geq 0}$, and we have defined $\beta_\pm \equiv 2\pm \sqrt{3}$. 
One can verify that for any $n_\pm$ and $k\in\Z_{\geq 0}$ we have $p_1\in\Z$ and crucially also 
$\mathtt{s}\in\Z$, where $\mathtt{s}$ is defined in \eqref{sdef}. 
Since the expression multiplying $(m_--m_+)$ on the right hand side of \eqref{tcond1} 
is rational, we may then always choose the integer $(m_--m_+)$ so that $t\in\Z$.

\subsection{Uplifting to $D=11$}

Uplifting the $D=7$ solution to $D=11$ using \cite{Cvetic:2000ah} (see appendix \ref{app:a}), we find that the metric is given by
\begin{align}\label{11duplift}
ds^2_{11}=\Delta^{1/3}ds^2_7+\Delta^{-2/3}\Big(e^{4\lambda_1+4\lambda_2}dw_0^2
&+e^{-2\lambda_1}[dw_1^2+w_1^2(d\chi_1-A^{12}_{(1)})^2]\nn
&+e^{-2\lambda_2}[dw_2^2+w_2^2(d\chi_2-A^{34}_{(1)})^2]\Big)\,,
\end{align}
where \begin{align}
\Delta=e^{-4\lambda_1-4\lambda_2}w_0^2+e^{2\lambda_1}w_1^2+e^{2\lambda_2}w_2^2\,.
\end{align}
Here $\Delta\chi_i=2\pi$ and $(w_0,w_1,w_2)$, satisfying $w_0^2+w_1^2+w_2^2=1$ and parametrising a quadrant of an $S^2$, together parametrise an $S^4$.
We can take, for example,
\begin{align}
w_0=\sin\xi,\qquad
w_1=\cos\xi\cos\theta,\qquad
w_2=\cos\xi\sin\theta,
\end{align}
with $-\pi/2\le\xi\le \pi/2$, $0\le\theta\le \pi/2$.

Uplifting the $AdS_3\times\spindle_1\ltimes\spindle_2$ solution \eqref{d7sssol}-\eqref{d7sssol2}, we see that the eight-dimensional internal space is an $S^4$ fibration over $\spindle_1\ltimes\spindle_2$. 
More precisely, here one can regard $S^4\subset \R\oplus\C\oplus\C$, with $w_0$ a coordinate 
on the first factor, and $(w_i,\chi_i)$ being polar coordinates on the two copies of $\C$, $i=1,2$. 
The two factors of $\C$ are then fibred over the seven-dimensional spacetime via the 
$U(1)$ gauge fields $A^{12}_{(1)}$, $A^{34}_{(1)}$, respectively. As such, this fibration 
is well-defined only if the periods of the corresponding gauge field fluxes 
$F^{12}_{(2)}\equiv dA^{12}_{(1)}$, $F^{34}_{(2)}\equiv dA^{34}_{(1)}$
are appropriately quantised through two-cycles in the base $AdS_3\times M_4$. 
We note first that equation \eqref{fluxcond} already implies that this 
is the case for a copy of the fibre $\spindle_2$ of $M_4$, 
and indeed this defines the twisting parameters $p_i\in\Z$, $i=1,2$. 
We next define the two-cycles $S_a\equiv \{y=y_a\}$,
$a=2,3$, to be the two sections defined by the two poles of the fibre $\spindle_2$.
From \eqref{d7sssol2} we then compute 
\begin{align}\label{fluxF12}
\frac{1}{2\pi}\int_{S_2} F^{12}_{(2)} & = \left(\frac{q_1}{h_1(y_2)}-1\right)\left(-\frac{4}{3}\frac{m_--m_+}{2m_-m_+}\right)\nonumber\\ &  = \frac{p_1 t[p_1+p_2+6(n_+-p_1)-\mathtt{s}]}{6m_-m_+n_-n_+(n_--n_+)}\, ,
\end{align}
where we have used \eqref{5dflux}, the results of section \ref{sec:2.1} and $t\in\mathbb{Z}$ was defined in
\eqref{tcond1}. Now  $y=y_2$ is the $\Z_{n_+}$ orbifold singularity of the fibre spindle $\spindle_2=\mathbb{WCP}^1_{[n_-,n_+]}$, 
while each $S_a\cong \spindle_1=\mathbb{WCP}^1_{[m_-,m_+]}$
for $a=2,3$ is a copy of the base spindle. 
Similar to \eqref{tqncond1}, we demand\footnote{It may well be possible to relax this condition and demand that
the flux is an integer multiple of $1/\mathrm{lcm}\{m_-,m_+,n_+\}$.} that the flux number in \eqref{fluxF12} 
is an integer divided by $(m_-m_+)$.
On the other hand $t\in\mathbb{Z}$ is determined by~\eqref{tcond1}. 

We may write down a family of solutions to these integrality constraints as follows. 
First introduce
\begin{align}\label{teesoln}
t = 6n_-n_+(n_--n_+)\uu\,,
\end{align}
where $\uu\in\Z$ is an \emph{arbitrary} integer. Then the condition \eqref{tcond1} reads
\begin{align}\label{mmp}
m_--m_+ = - [\mathtt{s}-(p_1+p_2)][\mathtt{s}+2(p_1+p_2)]\, \uu\, .
\end{align}
For the family \eqref{pkfamily}, recall that $\mathtt{s}\in\Z$, and the right hand side 
of \eqref{mmp} is manifestly an integer, as required, and moreover may be 
regarded as fixing $m_--m_+$ in terms of the arbitrary integers $n_\pm,k,\uu\in\Z$. 
It is then immediate from the expression in \eqref{fluxF12} that this 
flux number is an integer multiple of $1/m_-m_+$. 

One can then verify that the flux numbers for both $F^{12}_{(2)}$ and $F^{34}_{(2)}$ 
over the remaining two-cycles in $M_4$ are automatically quantised appropriately. For example, 
the flux of $F^{12}_{(2)}/2\pi$ through $S_3=\{y=y_3\}$ may be computed, 
with the expression  found to be consistent with the homology relation $S_3-S_2=\frac{t}{m_-m_+}\spindle_2\in H_2(M_4,\R)$.
On the other hand, the fluxes of $F^{34}_{(2)}$ are given by the same expressions 
as for $F^{12}_{(2)}$, but with $p_1$ and $p_2$ exchanged,  where recall the latter are constrained 
to obey $p_1+p_2=n_-+n_+$. 

Next we recall that the R-symmetry gauge field flux is
$F^R_{(2)}\equiv F^{12}_{(2)}+F^{34}_{(2)}$. A computation then  gives
\begin{align}\label{FRS2}
\frac{1}{2\pi}\int_{S_2}F^{R}_{(2)} & =\left[-\frac{1}{n_+}-\frac{[\mathtt{s}-(p_1+p_2)][\mathtt{s}+2(p_1+p_2)]}{6n_-n_+(n_--n_+)}\right]\frac{t}{m_-m_+}\nonumber\\
& = -\frac{1}{n_+}\frac{t}{m_-m_+}+\frac{m_--m_+}{m_-m_+}\, ,
\end{align}
and similarly we find
\begin{align}\label{FRS3}
\frac{1}{2\pi}\int_{S_3}F^{R}_{(2)}& =  \frac{1}{n_-}\frac{t}{m_-m_+}+\frac{m_--m_+}{m_-m_+}\, .
\end{align}
From \eqref{teesoln} we see that $n_\pm $ divide $t$ and hence the fluxes in
\eqref{FRS2}, \eqref{FRS3} are indeed integers divided by $(m_- m_+)$ as we demanded above.
Also, via the homology relation $S_3-S_2=\frac{t}{m_-m_+}\spindle_2$, the above equations immediately confirm
\begin{align}
\int_{\spindle_2} F^R_{(2)} = \frac{1}{n_-}+\frac{1}{n_+}= \int_{\spindle_2}c_1(\spindle_2)\, ,
\end{align}
consistent with \eqref{fluxcond}. Recall that the $S^4$ bundle over $M_4$ 
is twisted via embedding $S^4\subset \R\oplus \C\oplus\C$, where the 
gauge fields $A^{12}_{(1)}$, $A^{34}_{(1)}$ fibre 
the two copies of $\C$, respectively. If the total space of the corresponding $\C^2=\C\oplus \C$ bundle over $M_4$ were Calabi-Yau, the fluxes 
in \eqref{FRS2}, \eqref{FRS3} would agree with  the 
first Chern class $c_1(M_4)$ of $M_4$, integrated through the two sections $S_a$. 
At the section $S_2\cong\mathbb{WCP}^1_{[m_-,m_+]}$, the tangent bundle of $M_4$ splits into a direct sum, where the 
complex tangent bundle to the section is simply $\mathcal{O}(m_-+m_+)$, 
with Chern number  $\frac{m_-+m_+}{m_-m_+}$, 
 while the normal bundle is 
$\mathcal{O}(-t)$, with Chern number $-\frac{t}{n_+m_-m_+}$. 
Notice here that the normal direction is a $\Z_{n_+}$ singularity,
hence the extra factor of $n_+$.
 One can see precisely this structure in \eqref{FRS2},  except we have $m_--m_+$ rather than $m_-+m_+$. 
 Similar remarks apply to the section $S_3$, where the normal bundle is instead $\mathcal{O}(t)$, which is a 
 $\Z_{n_-}$ singularity. 
 Because of this, the total space of the $\C^2$ bundle is \emph{not} Calabi-Yau, 
but \emph{only} due to the relative minus sign in the $m_--m_+$ terms in \eqref{FRS2}, \eqref{FRS3}. 
This may have been anticipated,
since the original twist over the $\spindle_1$ spindle is an anti-twist, 
which is reflected in the above formulae.

Having ensured that the $D=11$ spacetime is a well-defined orbifold, we may next turn to the 
four-form flux. There are two natural four-cycles: 
fixing a point on the base $M_4=\spindle_1\ltimes\spindle_2$ we obtain a copy of the fibre $S^4$. 
On the other hand, if we fix either the north or south pole section $w_0=\pm1 $ of $S^4$, we obtain 
copies of the base  $M_4=\spindle_1\ltimes\spindle_2$. The four-form flux of the $D=11$ solution has several terms which can be obtained from appendix \ref{app:a}, which
we won't write explicitly here. Instead we focus on the terms that are relevant for quantising the four-form flux through the above cycles.
Specifically we have
\begin{align}\label{efffour}
F_{(4)}
&=\frac{w_1w_2}{w_0}U\Delta^{-2}dw_1\wedge dw_2\wedge
(d\chi_1-A^{12}_{(1)})\wedge (d\chi_2-A^{34}_{(1)}) \nn
&\qquad - w_0\frac{1}{3}F_{(2)}\wedge dy\wedge d{\yphi}+\dots \, ,
\end{align}
where
\begin{align}
U&=\left(e^{-8\lambda_1-8\lambda_2}-2e^{-2\lambda_1-4\lambda_2}-2e^{-4\lambda_1-2\lambda_2}\right)w_0^2\nn
&\quad -\left(e^{-2\lambda_1-4\lambda_2}+2e^{2\lambda_1+2\lambda_2}\right)w_1^2 -\left(e^{-4\lambda_1-2\lambda_2}+2e^{2\lambda_1+2\lambda_2}\right)w_2^2\,,
\end{align}
and the last term in $F_{(4)}$ in \eqref{efffour} arises from a $D=7$ contribution involving $*_7S^5_{(3)}$ which contains
the $D=5$ field strength $F_{(2)}$, as one sees from \eqref{aessckk}.

To carry out flux quantisation we first rescale the metric by $L^2$ and the four-from by $L^3$ (which does indeed give another $D=11$ solution).
We can then integrate the first term in \eqref{efffour} on the $S^4$ at a fixed point on $M_4=\spindle_1\ltimes\spindle_2$ 
and one finds 
\begin{align}
\frac{1}{(2\pi\ell_p)^3}\int_{S^4}F_{(4)}=\frac{L^3}{\pi \ell_p^3}\equiv N\,,
\end{align}
where $N$ is interpreted as the number of M5-branes wrapping $M_4\equiv \spindle_1\ltimes\spindle_2$. We can also integrate the four-form flux \eqref{efffour} along the orbifold four-cycle $M_4$ in the $D=7$ solution.
Representatives $M_4^\pm$ for this cycle (with opposite orientation) are obtained at the north or south pole of the $S^4$ fibre $w_0=\pm 1$ and we find 
\begin{align}\label{other4form}
\frac{1}{(2\pi\ell_p)^3}\int_{M_4^\pm}F_{(4)}=\pm\frac{m_--m_+}{m_- m_+}\frac{p_1 p_2}{n_- n_+(2(n_-+n_+)+\mathtt{s})}N\,.
\end{align}
Since the total $D=11$ spacetime has orbifold singularities, it is not clear what precise quantisation condition 
should be imposed on this flux. For the family of solutions we have discussed, with 
$p_1$ given by \eqref{pkfamily}, the expression \eqref{other4form} is rational but not in general integer. 
Of course, by choosing $N$ appropriately, it can be made integer. 

Finally,  we can also calculate the central charge of the dual $d=2$, $\mathcal{N}=(0,2)$ SCFT. The  $D=7$ Newton's constant 
is given by $(G_{(7)})^{-1}=N^3/(6\pi^2)$ (see, for example, appendix A.3 of \cite{Ferrero:2021etw}).
We then obtain the $D=3$ Newton's constant by reducing the $D=7$ theory on $M_4=\spindle_1\ltimes \spindle_2$ to get
\begin{align}
(G_{(3)})^{-1}&=(G_{(7)})^{-1}\int dx d \psi dy d\phi\,\left(\frac{1}{18}y\right)\nn
&=(G_{(7)})^{-1}\Delta x \Delta \psi\frac{1}{36}(y_3^2-y_2^2)\Delta \phi\,.
\end{align}
Thus  the $d=2$ central charge $c=(3/2)(G_{(3)})^{-1}$ can be written in the form
\begin{align}\label{finalcc}
c=\frac{4(m_--m_+)^3}{3m_- m_+(m_-^2+m_- m_+ +m_+^2)}a_{4d}\,,
\end{align}
where $a_{4d}$ is given in \eqref{sugracentralcharge},  where the latter is rational for the family 
of solutions we have discussed.

\section{$AdS_3\times  \Sigma_g\ltimes \spindle_2$ solutions}
Minimal $D=5$ gauged supergravity also admits a supersymmetric 
$AdS_3\times  H_2$ solution, where $H_2$ is a constant curvature metric on hyperbolic space.
After taking a discrete quotient we get an $AdS_3\times  \Sigma_g$ solution, where $\Sigma_g$ is a Riemann surface
with genus $g>1$. This solution is again dual to a $d=2$ SCFT with $\mathcal{N}=(0,2)$ supersymmetry. 
This is the standard topological twist solution and arises as the near-horizon limit of a black string solution 
\cite{Naka:2002jz,Klemm:2000nj}.

Using the consistent truncation ansatz  \eqref{const1}-\eqref{const3} we can also uplift this solution on the spindle $\spindle_2$ to obtain an 
$AdS_3\times  \Sigma_g\ltimes \spindle_2$ solution of $D=7$ gauged gravity with $\spindle_2$ non-trivially fibred over $\Sigma_g$.
The $D=7$ metric is given by
\begin{align}
\label{eq:D=7_Riemann_spindle_1}
ds_7^2&=\frac{4(yP)^{1/5}}{9}\left[ds^2(AdS_3)+\frac{3}{4}ds^2(\Sigma_g)+\frac{9y}{16Q}dy^2+\frac{9Q}{4P}\left(d\phi-\frac{2}{3}{\omega}\right)^2\right]\,.
\end{align}
Here $ds^2(\Sigma_g)$ is normalised so that the Ricci scalar is $R(\Sigma_g)=-2$ and $\omega$ is the Levi-Civita connection
one-form satisfying $d\omega=-\text{vol}(\Sigma_g)$. Thus the volume of the Riemann surface is $\int_{\Sigma_g}\vol(\Sigma_g)=4\pi(g-1)$.
The remaining fields take the form
\begin{align}
\label{eq:D=7_Riemann_spindle_2}
A^{12}_{(1)}&=\left(\frac{q_1}{h_1}-1\right)\left(d\phi-\frac{2}{3}{\omega}\right)\,,\nn
A^{34}_{(1)}&=\left(\frac{q_2}{h_2}-1\right)\left(d\phi-\frac{2}{3}{\omega}\right)\,,\nn
S^5_{(3)}&=\frac{8y}{27}\text{vol}(AdS_3)-\frac{2yQ}{3h_1h_2}d\phi\wedge \text{vol}(\Sigma_g)\,,\nn
e^{2\lambda_i}&=\frac{(yP)^{2/5}}{{h_i}}\,.
\end{align}

Notice that the spindle $\spindle_2$ is non-trivially fibred over $\Sigma_g$. 
To ensure that this fibration is well-defined (in the orbifold sense), we demand that the
one-form determining the fibration, $\eta\equiv \frac{2\pi}{\Delta\phi}(d\phi-\frac{2}{3}{\omega})$, is globally defined. This
requires that 
\begin{align}\label{tqncond}
\frac{1}{2\pi} \int_{\Sigma_g}d\eta= {{t}}\, ,\qquad {t}\in \mathbb{Z}\,,
\end{align}
and hence we need to impose the quantisation condition relating the spindle quantum numbers with the genus:
\begin{align}\label{tcond}
 {t}={12}(g-1)\frac{n_-\,n_+\,(n_--n_+)}{[\mathtt{s}-(p_1+p_2)]\,[\mathtt{s}+2(p_1+p_2)]}
 \in \mathbb{Z}\,.
\end{align}
The global discussion of the resulting space $M_4=\Sigma_g\ltimes\spindle_2$ is then 
very similar to that in the previous section. Specifically, one 
begins with the complex line bundle $\mathcal{O}(t)$ over the Riemann 
surface $\Sigma_g$, and then uses the $U(1)$ transition function 
for this bundle to construct the associated $\spindle_2$ fibration over 
$\Sigma_g$, with $U(1)$ acting on $\spindle_2$ by azimuthal rotations around the poles. 
The resulting space is then a spindly version of a rationally ruled surface, replacing 
the $\mathbb{CP}^1$ fibres by $\spindle_2=\mathbb{WCP}^1_{[n_-,n_+]}$. 

The twisting parameter $t\in\Z$ is constrained to satisfy \eqref{tcond}. Similar to the
previous section we can solve this by first writing 
\begin{align}\label{tsubu}
t = 12n_-n_+(n_--n_+)\uu\, ,
\end{align} 
with $\uu\in\Z$ arbitrary, and then imposing that the fibre data for $\spindle_2$ is 
given by the family of solutions in \eqref{pkfamily}. One then chooses the genus $g$ 
to be
\begin{align}
g= 1 + [\mathtt{s}-(p_1+p_2)]\,[\mathtt{s}+2(p_1+p_2)]\uu\, ,
\end{align}
where the right hand side is now manifestly an integer. This family of solutions 
is then specified by $n_\pm, k,\uu\in\Z$, where recall that for this family also $\mathtt{s}\in\Z$. 

We can now uplift on $S^4$ to obtain a $D=11$ solution exactly as in the previous section. 
There are again two sections $S_a=\{y=y_a\}\cong \Sigma_g$, $a=2,3$, and we compute
\begin{align}
\frac{1}{2\pi}\int_{S_2} F^{12}_{(2)} & = \left(\frac{q_1}{h_1(y_2)}-1\right)\frac{4}{3}(g-1)  = 
\frac{p_1 t[p_1+p_2+6(n_+-p_1)-\mathtt{s}]}{6n_-n_+(n_--n_+)}\, ,
\end{align}
similarly to \eqref{fluxF12}. Substituting for $t\in\Z$ using \eqref{tsubu}, this 
flux number is an integer for the family of solutions described above. 
The remaining flux numbers for $F^{12}_{(2)}$, $F^{34}_{(2)}$ are similarly 
integers.  However, a key difference with the $\spindle_1\ltimes \spindle_2$ solutions in the previous 
section is that the $\C^2$ fibration over $M_4=\Sigma_g\ltimes\spindle_2$ is now a
Calabi-Yau four-fold. Essentially, this is because we have doubly uplifted two 
twist solutions. To see this, we compute the R-symmetry fluxes
\begin{align}
\frac{1}{2\pi}\int_{S_2}F^R_{(2)} = -\frac{1}{n_+}t - 2(g-1)\, , \quad 
\frac{1}{2\pi}\int_{S_3}F^R_{(2)} = \frac{1}{n_-}t - 2(g-1)\, ,
\end{align}
which are both integers since $n_\pm $ divide $t$ from \eqref{tsubu}.
On the other hand, these expressions are precisely 
$c_1(M_4)$ integrated over the cycles $S_2$, $S_3$, respectively, 
where recall that $\int_{\Sigma_g}c_1(T\Sigma_g)=-2(g-1)$, and 
the normal bundles of the cycles are respectively $\mathcal{O}(-t)$ and 
$\mathcal{O}(t)$. 
This implies that the $\C^2$ bundle over $M_4$ has
zero first Chern class, making the total space a Calabi-Yau four-fold. 

We normalise the solution
so that there are $N$ units of quantised four-form flux through the $S^4$ fibre. There is then also the flux through the four-cycles 
$M_4^\pm \cong \Sigma_g\ltimes \spindle_2$ at the north and south pole sections $w_0=\pm 1$ of the $S^4$, where we compute
\begin{align}\label{fluxother}
\frac{1}{(2\pi\ell_p)^3}\int_{M_4^\pm}F_{(4)}=\mp2(g-1)\frac{p_1 p_2}{n_- n_+(2(n_-+n_+)+\mathtt{s})}N\, .
\end{align}
As in the previous section, in general this is rational for the above family, but by choosing $N$ appropriately 
one can ensure that the fluxes are  integer. 

Finally, the central charge of the $d=2$, $\mathcal{N}=(0,2)$ SCFT can be computed in exactly the same was the previous section, 
and is given in the large $N$ limit by
\begin{align}\label{ftcctopt}
c=\frac{32}{3}(g-1)a_{4d}\,,
\end{align}
where $a_{4d}$ is given in \eqref{sugracentralcharge}. 
This result for the central charge is in agreement with the general field theory result
of \cite{Benini:2015bwz} for general $d=4$, $\mathcal{N}=1$ SCFTs reduced on a Riemann surface with a topological twist.

\section{Field theory}
We can calculate the central charge of the $d=2$, $\mathcal{N}=(0,2)$ SCFTs dual to the $AdS_3$ solutions discussed in the previous two sections using field theory arguments in a two step process. We start with the $d=6$, $\mathcal{N}=(0,2)$
SCFT living on a stack of $N$ M5-branes. We reduce this $d=6$ SCFT on a spindle $\spindle_2$ with magnetic fluxes in the twist class 
to get a $d=4$, $\mathcal{N}=1$ SCFT. From the results of \cite{Ferrero:2021wvk}, based on studying the anomaly polynomial for the M5-brane and using a-maximisation,
the central charge of the $d=4$ SCFT is given, in the large $N$ limit, by 
\begin{align}
a_{4d}=\frac{3p_1^2 p_2^2(p_1+p_2+\mathtt{s})}{8n_- n_+ (n_--p_1)(p_2-n_-)(\mathtt{s}+2(p_1+p_2))^2}N^3\,,
\end{align}
exactly as can be derived from the dual gravity solution \eqref{sugracentralcharge}.

We first consider reducing this $d=4$, $\mathcal{N}=1$ SCFT on the spindle $\spindle_1$, specified by co-prime integers ${m_+,m_-}$ and
with a magnetic flux in the anti-twist class, to get a $d=2$, $\mathcal{N}=(0,2)$ SCFT. Using the results of \cite{Ferrero:2020laf}, based on the anomaly polynomial of a general $d=4$, $\mathcal{N}=1$ SCFT, the central charge of the $d=2$ SCFT is given in the large $N$ limit by
\begin{align}
c=\frac{4(m_--m_+)^3}{3m_- m_+(m_-^2+m_- m_+ +m_+^2)}a_{4d}\,,
\end{align}
exactly as we derived from the gravity calculation in \eqref{finalcc}.

We can also carry out a similar analysis after reducing the $d=4$, $\mathcal{N}=1$ SCFT on a Riemann surface $\Sigma_g$ with 
a topological twist. In fact this is an example of a ``universal twist" and we can immediately use the results
of \cite{Benini:2015bwz} which gives the central charge in the large $N$ limit as
\begin{align}
c=\frac{32}{3}(g-1)a_{4d}\,,
\end{align}
again exactly as we derived from the gravity calculation in \eqref{ftcctopt}.

It is also possible to derive these field theory results in a one-step process, by directly reducing the anomaly polynomial for the M5-brane
on the orbifold four-cycles. 
For example, in
compactifying the $d=6$ theory on $\spindle_1\ltimes\spindle_2$ we need to take into account that the R-symmetry of the
$d=2$ SCFT arises from a mixture of an R-symmetry of the $d=6$ SCFT with the $U(1)\times U(1)$
global symmetry that arises from the isometries of $\spindle_1\ltimes\spindle_2$. 
In appendix \ref{app:c} we carry out this analysis, which generalises the analysis of
\cite{Ferrero:2020laf,Ferrero:2021wvk} on individual spindles. While the final answer is identical, we have included some details
because such an analysis would be needed in compactifying the $d=6$ theory on more general orbifolds. 

\section{Discussion}\label{sec:discussion}

In this paper we have found two new families of $AdS_3$ solutions of $D=11$ supergravity, describing M5-branes wrapped on four-dimensional orbifolds $M_4$. 
In both cases $M_4$ takes the form of a spindle $\spindle_2$ fibred over another two-dimensional space: either another spindle $\spindle_1$, or a 
smooth Riemann surface $\Sigma_g$ of genus $g>1$. The  solutions are dual to $d=2$, $\mathcal{N}=(0,2)$ SCFTs, 
and a computation of the central charges of these theories using anomaly polynomials perfectly matches the supergravity results.
In the case of $M_4=\Sigma_g\ltimes \spindle_2$, the solutions can be naturally interpreted as M5-branes wrapping 
an orbifold four-cycle, which is holomorphically embedded inside a Calabi-Yau four-fold, generalising \cite{Gauntlett:2000ng,Benini:2013cda}.
Such an interpretation is not available for the solutions with $M_4=\spindle_1\ltimes \spindle_2$; this feature,
which has an analogue for all of the solutions involving spindles with an anti-twist that have been constructed, deserves a better understanding.

A key ingredient in the construction is a new consistent Kaluza-Klein truncation of $D=7$ gauged supergravity on a spindle 
down to $D=5$ minimal gauged supergravity. The solutions have then been obtained by a double uplifting procedure, 
starting with $AdS_3\times \spindle_1$ or $AdS_3\times \Sigma_g$ solutions of minimal $D=5$ gauged supergravity, respectively, 
uplifting to $D=7$ on $\spindle_2$, and then uplifting again to $D=11$ on $S^4$. This consistent truncation is local in the gravity fields. 
In particular the analysis we have done will go through also for the (singular) half-spindle solutions of the general type studied in
\cite{Bah:2021mzw, Bah:2021hei,Couzens:2021tnv,Suh:2021ifj}. More generally, analogous consistent truncations can
also be carried out for other $AdS_d\times \spindle$ solutions, such as \cite{Faedo:2021nub},
leading to additional $\spindle\ltimes\spindle$ solutions.

The structure of the $D=7$ 
solutions with $M_4=\spindle_1\ltimes \spindle_2$ that we have found is quite remarkable: the solutions are co-homogeneity two, with the various supergravity fields depending non-trivially on 
two coordinates $x$ and $y$, but they also exhibit a remarkable separation of variables.  Such a separation of variables 
in solutions to the Einstein equations is often associated with the existence of a Killing (or Killing-Yano) tensor, and it would be interesting 
to investigate this further. Indeed, it would have been extremely difficult to find the $D=7$ solutions directly, without 
any {\it a priori} understanding of how to separate variables in this way, and there may be similar classes of solutions that generalise those we have 
found, also in other supergravity theories. We note that the corresponding uplifted $D=11$ $AdS_3$ solutions 
are different to those found in \cite{Gauntlett:2006qw}. This naturally begs the question of how 
the solutions fit into a $G$-structure classification, extending  \cite{Gauntlett:2006ux}.

Both families of supergravity solutions depend on a number of integer parameters, and we expect there to be more general solutions 
of this type. For example, one might anticipate solutions with $M_4=\spindle_1\ltimes\spindle_2$, with arbitrary spindle 
data $m_\pm$, $n_\pm$, for $\spindle_1$, $\spindle_2$, respectively, with an arbitrary twisting parameter $t\in\Z$ 
describing the fibration, and where the $S^4\subset\R\oplus\C^2$ fibration is specified by two further integer Chern numbers. 
This is a seven-parameter family, while the solutions we have found have only five parameters; presumably this is due to 
the particular way we have constructed them as a double uplift. 
The larger conjectured  family would also include \emph{smooth} $M_4$: 
setting $m_\pm=1=n_\pm$ gives Hirzebruch surfaces 
$M_4=\mathbb{F}_{|t|}$.

The results we have presented open the door for potentially many more 
new orbifold solutions, also in other supergravity theories in different dimensions. 
This raises a key question that we have left open in this paper: what is the appropriate 
four-form flux quantisation condition in M-theory when the $D=11$ spacetime 
has orbifold singularities? One approach to this would be to resolve (at least topologically) 
the singularities, quantise the flux on this smooth resolution, and then take the  singular limit. 
This would lead to \emph{rationally} quantised flux, and we have shown this is always possible 
to impose for the solutions we have presented, by appropriately choosing the parameters, but the precise
quantisation condition required is not yet clear. We leave this, and many of the other interesting questions 
we have raised above, for future work.

\section*{Acknowledgements}

\noindent 
We thank Rahim Leung, Achilleas Passias, Alessandro Tomasiello for helpful discussions and Dario Martelli for collaboration on related topics.
This work was supported in part by STFC grants  ST/T000791/1 and 
ST/T000864/1.  KCMC is supported by an Imperial College President’s PhD Scholarship and JHTF is supported by an STFC PhD studentship.
JPG is supported as a Visiting Fellow at the Perimeter Institute. 

\appendix

\section{$D=7$ gauged supergravity}\label{app:a}

\subsection{Maximal $D=7$ gauged supergravity}
The $D=7$ maximal $SO(5)$ gauged supergravity \cite{Pernici:1984xx} can be obtained by performing a consistent reduction of $D=11$ supergravity on $S^4$. The details of this truncation, including the explicit demonstration of its consistency, are given in \cite{Nastase:1999cb,Nastase:1999kf}. The bosonic field content of the theory is comprised of a metric, $SO(5)$ Yang-Mills one-forms $A^{ij}_{(1)}$ transforming in the $\boldsymbol{10}$ of $SO(5)$, three-forms $S^i_{(3)}$ transforming in the $\boldsymbol{5}$ of $SO(5)$, and fourteen scalar fields given by a symmetric unimodular matrix $T_{ij}$ that parametrises the coset $SL(5,\mathbb{R})/SO(5)$. In the notation of \cite{Cvetic:2000ah}, the Lagrangian of the bosonic sector of the theory is given by 
\begin{align}\label{lag7}
\mathcal{L}_{(7)}=\,&(\textit{R}-V) \vol_7-\frac{1}{4}T^{-1}_{ij}{\ast_7 D}T_{jk}\wedge T^{-1}_{kl} DT_{li}-\frac{1}{4}T^{-1}_{ik}T^{-1}_{jl} {\ast_7 F^{ij}_{(2)}}\wedge F^{kl}_{(2)}+\frac{1}{g_c}\Omega_{(7)}
\nn
&
-\frac{1}{2}T_{ij}\, {\ast_7 S^i_{(3)}}\wedge S^j_{(3)}+\frac{1}{2g_c}S^i_{(3)}\wedge DS^i_{(3)}-\frac{1}{8g_c}\epsilon_{ij_1j_2j_3j_4}\,S^i_{(3)}\wedge F^{j_1j_2}_{(2)}\wedge F^{j_3j_4}_{(2)}\,,
\end{align}
where
\begin{equation}
\begin{split}
DT_{ij}&\equiv dT_{ij}+g_cA^{ik}_{(1)}T_{kj}+g_cA^{jk}_{(1)}T_{ik}\,,\\
DS^i_{(3)}&\equiv dS^i_{(3)}+g_cA^{ij}_{(1)}\wedge S^j_{(3)}\,,\\
F^{ij}_{(2)}&\equiv dA^{ij}_{(1)}+g_cA^{ik}_{(1)}\wedge A^{kj}_{(1)}\,,
\end{split}
\end{equation}
the scalar potential is given by
\begin{equation}
V=\frac{1}{2}g_c^2\bigg[2\mathrm{Tr}(T^2)-(\mathrm{Tr}(T))^2\bigg]\,,
\end{equation}
and the variation of the Chern-Simons term with respect to $A^{ij}_{(1)}$ is given by
\begin{equation}
\delta \Omega_{(7)}=\frac{3}{4}\delta^{j_1j_2j_3j_4}_{i_1i_2kl} \, F^{i_1i_2}_{(2)}\wedge F^{j_1j_2}_{(2)}\wedge F^{j_3j_4}_{(2)}\wedge \delta A^{kl}_{(1)}\,.
\end{equation}
The explicit equations of motion can be found in appendix A of \cite{MatthewCheung:2019ehr} (which has corrected a typo in the scalar equation of motion in \cite{Cvetic:2000ah}).
As evident from \eqref{lag7}, we use mostly plus signature and also $\epsilon_{12345}=+1$.
In the appendices we have included factors of $g_c$ and in the main text we have set $g_c=1$ for clarity.

Any solution of maximal $D = 7$ gauged supergravity can be uplifted on $S^4$ to obtain a solution of $D = 11$ supergravity \cite{Nastase:1999cb,Nastase:1999kf,Cvetic:2000ah}. Following \cite{Cvetic:2000ah}, the $D = 11$ metric is given by
\begin{align}\label{eq:general_11d_metric_uplift}
ds^2_{11}=\Delta^{1/3}ds^2_7+\frac{1}{g_c^2}\Delta^{-2/3}T^{-1}_{ij}D\mu^iD\mu^j\,,
\end{align}
where $\mu^i$ are constrained coordinates on $S^4$ satisfying $\mu^i\mu^i= 1$, and
\begin{align}
\Delta=T_{ij}\mu^i\mu^j\,,\quad D\mu^i=d\mu^i+g_cA^{ij}_{(1)}\mu^j\,.
\end{align}
The $D=11$ four-form flux is given by
\begin{align}
\label{eq:general_four_form_uplift}
F_{(4)}=&\frac{1}{4!}\epsilon_{i_1\cdots i_5}\Big[-\frac{1}{g_c^3}U \Delta^{-2}\mu^{i_1}D\mu^{i_2}\wedge\cdots\wedge D\mu^{i_5}\nn
&+\frac{4}{g_c^3}\Delta^{-2} T^{i_1m}DT^{i_2n}\mu^m\mu^n\wedge D\mu^{i_3}\wedge D\mu^{i_4}\wedge D\mu^{i_5}\nn
&+\frac{6}{g_c^2}\Delta^{-1} F^{i_1i_2}_{(2)}\wedge D\mu^{i_3}\wedge D\mu^{i_4}T^{i_5j}\mu^j\Big]-T_{ij}{\ast_7 S^i_{(3)}}\mu^j+\frac{1}{g_c}S^i_{(3)}\wedge D\mu^i\,, 
\end{align}
where
\begin{align}
\begin{split}
U=2T_{ij}T_{jk}\mu^i\mu^k-\Delta T_{ii}\,.
\end{split}
\end{align}
The uplifted solution will then solve the following $D=11$ equations of motion
\begin{align}
R_{MN}&=\frac{1}{12}\left(F_{MPQR}F_N^{\phantom{N}PQR}-\frac{1}{12}g_{MN}F^2\right)\,,\nn
d{\ast_{11} F_{(4)}}&=\frac{1}{2}F_{(4)}\wedge F_{(4)}\,.
\end{align}
We also note that the $AdS_7$ vacuum solution with $A^{ij}_{(1)}=S^i_{(3)}=0$ and $T_{ij}=\delta_{ij}$ uplifts to the maximally supersymmetric $AdS_7\times S^4$ vacuum solution, with the radius of the $AdS_7$ given by $L_{AdS_7}={2}g_c^{-1}$ and the radius of the $S^4$ given by $g_c^{-1}$.

\subsection{The $U(1)\times U(1)$ truncation}
One can consistently truncate this to a $D=7$ $U(1)^2$ theory, as first discussed in \cite{Liu:1999ai}, by 
keeping just the two $U(1)$'s, $F^{12}_{(2)}$, $F^{34}_{(2)}$, one of the three-forms, $S^{5}_{(3)}$, and two scalars
\begin{align}
\begin{split}
T_{ij}=&\text{diag}\left(e^{2\lambda_1},e^{2\lambda_1},e^{2\lambda_2},e^{2\lambda_2},e^{-4\lambda_1-4\lambda_2}\right)\,.
\end{split}
\end{align}
The Lagrangian \eqref{lag7} then reduces to 
\begin{align}\label{eq:7d_U(1)2_lagrangian}
\mathcal{L}_{(7)}=&\,(\textit{R} -V)\text{vol}_7-6{\ast_7 d\lambda_1}\wedge d\lambda_1-6{\ast_7 d\lambda_2}\wedge d\lambda_2-8{\ast_7 d\lambda_1}\wedge d\lambda_2\nn
&-\frac{1}{2}e^{-4\lambda_1}{\ast_7 F^{12}_{(2)}}\wedge F^{12}_{(2)}-\frac{1}{2}e^{-4\lambda_2}{\ast_7 F^{34}_{(2)}}\wedge F^{34}_{(2)}-\frac{1}{2}e^{-4\lambda_1-4\lambda_2}{\ast_7 S^5_{(3)}}\wedge S^5_{(3)}\nn
&+\frac{1}{2g_c}S^5_{(3)}\wedge dS^5_{(3)}-\frac{1}{g_c}S^5_{(3)}\wedge F^{12}_{(2)}\wedge F^{34}_{(2)}+\frac{1}{2 g_c}A^{12}_{(1)}\wedge F^{12}_{(2)}\wedge  F^{34}_{(2)}\wedge  F^{34}_{(2)}\,,
\end{align}
where
\begin{align}
V=g_c^2\Big[\frac{1}{2} e^{-8 (\lambda_1+\lambda_2)}-4 e^{2 (\lambda_1+\lambda_2)}-2 e^{-2 (2 \lambda_1+\lambda_2)}-2 e^{-2 (\lambda_1+2 \lambda_2)}\Big]\,.
\end{align}
For configurations with $F^{12}_{(2)}\wedge F^{34}_{(2)}=0$, we can further consistently set $S^5_{(3)}=0$.

The equations of motion for the $D=7$ $U(1)^2$ gauged supergravity arising from \eqref{eq:7d_U(1)2_lagrangian} 
can be written in the form 
\begin{align}\label{pgqeom}
\mathcal{P}_1  &\equiv 3d{\ast_7 d\lambda_1}+2d{\ast_7 d \lambda_2}+\frac{1}{2}e^{-4\lambda_1}{\ast_7 F_{(2)}^{12}}\wedge F_{(2)}^{12}+\frac{1}{2}e^{-4\lambda_1-4\lambda_2}{\ast_7 S_{(3)}^5}\wedge S^5_{(3)}\nn
&\quad -g_c^2\left(2e^{-2(2\lambda_1+\lambda_2)}+e^{-2(\lambda_1+2\lambda_2)}-e^{-8(\lambda_1+\lambda_2)}-2e^{2(\lambda_1+\lambda_2)}\right){\ast_7\mathds{1}}=0\,,\nn
\mathcal{P}_2 &\equiv 2d{\ast_7 d\lambda_1}+3d{\ast_7 d \lambda_2}+\frac{1}{2}e^{-4\lambda_2}{\ast_7 F_{(2)}^{34}}\wedge F_{(2)}^{34}+\frac{1}{2}e^{-4\lambda_1-4\lambda_2}{\ast_7 S^5_{(3)}}\wedge S^5_{(3)}\nn
&\quad -g_c^2\left(e^{-2(2\lambda_1+\lambda_2)}+2e^{-2(\lambda_1+2\lambda_2)}-e^{-8(\lambda_1+\lambda_2)}-2e^{2(\lambda_1+\lambda_2)}\right){\ast_7\mathds{1}}=0\,,\nn
\mathcal{G}_1&\equiv d\left(e^{-4\lambda_1}{\ast_7 F_{(2)}^{12}}\right)
+e^{-4\lambda_1-4\lambda_2}{\ast_7S^5_{(3)}}\wedge F^{34}_{(2)}
=0\,,\nn
\mathcal{G}_2&\equiv d\left(e^{-4\lambda_2}{\ast_7 F_{(2)}^{34}}\right)
+e^{-4\lambda_1-4\lambda_2}{\ast_7S^5_{(3)}}\wedge F^{12}_{(2)}
=0\,,\nn
\mathcal{T}&\equiv dS^5_{(3)}-g_c e^{-4\lambda_1-4\lambda_2}{\ast_7  S_{(3)}^5}-F_{(2)}^{12}\wedge F_{(2)}^{34}=0\,,
\end{align}
and
\begin{align}
\label{eq:U(1)2_einstein}
R_{\mu\nu}&=6\partial_\mu\lambda_1\partial_\nu\lambda_1+6\partial_\mu\lambda_2\partial_\nu\lambda_2+8\partial_{(\mu}\lambda_1\partial_{\nu)}\lambda_2+\frac{1}{5}g_{\mu\nu}V \nn
&\quad +\frac{1}{2}e^{-4\lambda_1}\big(F^{12}_{\mu\rho}F^{12\rho}_{\nu}-\frac{1}{10}g_{\mu\nu}F^{12}_{\rho\sigma}F^{12\rho\sigma}\big)
+\frac{1}{2}e^{-4\lambda_2}\big(F^{34}_{\mu\rho}F^{34\rho}_{\nu}-\frac{1}{10}g_{\mu\nu}F^{34}_{\rho\sigma}F^{34\rho\sigma}\big)
\nn
&\quad +\frac{1}{4}e^{-4\lambda_1-4\lambda_2}\big(S^{5}_{\mu\rho\sigma}S^{5\rho\sigma}_{\nu}-\frac{2}{15}g_{\mu\nu}
S^{5}_{\rho\sigma\delta}S^{5\rho\sigma\delta}\big)\,.
\end{align}

To uplift solutions to $D=11$ on $S^4$ in the $U(1)^2$ truncation it is convenient to parametrise the four-sphere $S^4$ by writing
the constrained coordinates $\mu^i$ as
\begin{align}
\mu^{1}+i\mu^2=\cos\xi\cos{\theta}\,e^{i\chi_1}\,,\quad \mu^{3}+i\mu^4=\cos\xi\sin{\theta}\,e^{i\chi_2}\,, \quad \mu^5=\sin\xi\,,
\end{align}
with $-\pi/2\leq \xi \leq \pi/2$, $0\leq \theta \leq \pi/2$ and $0\leq \chi_1,\chi_2\leq 2\pi$. Using the above parametrisation, we can easily write down the uplift ansatz for the $D=7$ $U(1)^2$ theory
\begin{align}
\label{eq:11d_metric_uplift}
ds^2_{11}&=\Delta^{1/3}ds^2_7+\frac{1}{g_c^2}\Delta^{-2/3}\Big\{e^{4\lambda_1+4\lambda_2}dw_0^2+e^{-2\lambda_1}\left[dw_1^2+w_1^2\left(d\chi_1-g_cA^{12}_{(1)}\right)^2\right]\nn
&\qquad\qquad\phantom{=\Delta^{1/3}ds^2_7+\frac{1}{g_c^2}\Delta^{-2/3}[}+e^{-2\lambda_2}\left[dw_2^2+w_2^2\left(d\chi_2-g_cA^{34}_{(1)}\right)^2\right]\Big\}\,,
\end{align}
with
\begin{align}
\begin{split}
\Delta=e^{-4\lambda_1-4\lambda_2}w_0^2+e^{2\lambda_1}w_1^2+e^{2\lambda_2}w_2^2\,,
\end{split}
\end{align}
and
\begin{align}
\begin{split}
w_0=\sin\xi\,,\quad w_1=\cos\xi\cos\theta\,,\quad w_2=\cos\xi\sin\theta\,,
\end{split}
\end{align}
satisfying $w_0^2+w_1^2+w_2^2=1$.

Within the $D=7$ $U(1)^2$ theory, the $D=11$ four-form flux can be written as
 \begin{align}
\begin{split}
F_{(4)}&=\frac{w_1w_2}{g_c^3w_0}U\Delta^{-2}\,dw_1\wedge dw_2\wedge\left(d\chi_1-g_cA^{12}_{(1)}\right)\wedge \left(d\chi_2-g_cA^{34}_{(1)}\right) \\
&\quad +\frac{2w_1^2w_2^2}{g_c^3}\Delta^{-2}e^{2\lambda_1+2\lambda_2}\left(d\lambda_1-d\lambda_2\right)\wedge  \left(d\chi_1-g_cA^{12}_{(1)}\right) \wedge  \left(d\chi_2-g_cA^{34}_{(1)}\right)\wedge dw_0\\
&\quad +\frac{2w_0w_1w_2}{g_c^3}\Delta^{-2}\left[e^{-2\lambda_1-4\lambda_2}w_1dw_2\wedge \left(3d\lambda_1+2d\lambda_2\right)-e^{-4\lambda_1-2\lambda_2}w_2dw_1\wedge \left(2d\lambda_1+3d\lambda_2\right)\right]\\
&\quad\, \phantom{+\frac{2w_0w_1w_2}{g_c^3}\Delta^{-2}[}\wedge \left(d\chi_1-g_cA^{12}_{(1)}\right) \wedge  \left(d\chi_2-g_cA^{34}_{(1)}\right) \\
&\quad +\frac{1}{g_c^2}\Delta^{-1}F^{12}_{(2)}\wedge \left[w_0w_2\,e^{-4\lambda_1-4\lambda_2}dw_2-w_2^2e^{2\lambda_2}dw_0\right]\wedge \left(d\chi_2-g_cA^{34}_{(1)}\right)\\
&\quad +\frac{1}{g_c^2}\Delta^{-1}F^{34}_{(2)}\wedge \left[w_0w_1\,e^{-4\lambda_1-4\lambda_2}dw_1-w_1^2e^{2\lambda_1}dw_0\right]\wedge \left(d\chi_1-g_cA^{12}_{(1)}\right)\\
&\quad -w_0e^{-4\lambda_1-4\lambda_2} {\ast_7 S_{(3)}^5}+\frac{1}{g_c}S^5_{(3)}\wedge dw_0\,,
\end{split}
\end{align}
with
\begin{align}
U&=\left(e^{-8\lambda_1-8\lambda_2}-2e^{-2\lambda_1-4\lambda_2}-2e^{-4\lambda_1-2\lambda_2}\right)w_0^2\nn
&\quad -\left(e^{-2\lambda_1-4\lambda_2}+2e^{2\lambda_1+2\lambda_2}\right)w_1^2 -\left(e^{-4\lambda_1-2\lambda_2}+2e^{2\lambda_1+2\lambda_2}\right)w_2^2\,.
\end{align}

We also note that if we integrate the $D=11$ four-form flux over the $S^4$ at an arbitrary point on the $D=7$ spacetime, we get
\begin{align}
\int_{S^4}F_{(4)}&=\int_{S^4}\frac{w_1w_2}{g_c^3w_0}U\Delta^{-2}\,dw_1\wedge dw_2\wedge d\chi_1\wedge d\chi_2 \nn
&=8\pi^2\,.
\end{align}
That is, the dependence in the integrand on $\lambda_i$ drops out of the definite integral.

\section{Supersymmetry of $D=7$ gauged supergravity}\label{app:b}
We write the supersymmetry transformations for bosonic configurations of maximal $SO(5)$ gauged supergravity associated with
the conventions for the bosonic sector of
\cite{Cvetic:2000ah} as
\begin{align}\label{d7susypope}
\delta\psi_\mu=&\Big[\nabla_\mu+\frac{1}{4}Q_{\mu ij}\Gamma^{ij}-\frac{g_c}{20}T\gamma_\mu+\frac{1}{80}(\gamma_\mu{}^{\nu\rho}-8\delta^\nu_\mu\gamma^\rho)\Gamma_{ij}
\Pi_A^i\Pi_B^jF_{(2)\nu\rho}^{AB}\nn
&-\frac{1}{60}(\gamma_\mu{}^{\nu\rho\sigma}-\frac{9}{2}\delta_\mu^\nu\gamma^{\rho\sigma})\Gamma^i(\Pi^{-1}){}^A_iS_{(3)A\nu\rho\sigma}\Big]\epsilon\,,\nn
\delta\chi_i=&\Big[-\frac{1}{2}\gamma^\mu\Gamma^jP_{\mu ij}
+\frac{g_c}{2}(T_{ij}-\frac{1}{5}T\delta_{ij})\Gamma^j+\frac{1}{32}\gamma^{\mu\nu}(\Gamma_{kl}\Gamma_i-\frac{1}{5}\Gamma_i\Gamma_{kl})\Pi_A^k\Pi_B^lF_{(2)\mu\nu}^{AB}\nn
&+\frac{1}{120}\gamma^{\mu\nu\rho}(\Gamma_i{}^j-4\delta^j_i)(\Pi^{-1}){}^A_jS_{(3)A\mu\nu\rho}\Big]\epsilon\,.
\end{align}
Our $D=7$ gamma matrices satisfy $\gamma_{0123456}=+1$ and $\Gamma_{12345}=+1$. 
Here $\Pi^k_A$, $A=1,\dots, 5$ also parametrise the coset $SL(5,\mathbb{R})/SO(5)$ and $T_{ij}=\Pi^{-1}{}_i^A\Pi^{-1}{}_j^B\delta_{AB}$. In addition, $P_{\mu ij}$ and $Q_{\mu ij}$ are defined as the symmetric and antisymmetric parts of
$\Pi^{-1}{}_i^A(\delta_A^B\partial_\mu+\frac{g_c}{2}A_{\mu A}{}^B)\Pi_B^k\delta_{kj}$, respectively.
We have determined these from the 
literature\footnote{We can make a comparison with the work of \cite{Gauntlett:2000ng}: if we make the redefinitions $g_c\to m$, $A\to 2B$, $S\to -(6m/\sqrt{3}) S$ as well as take $\gamma \to-\gamma^\mu$ then \eqref{lag7}, \eqref{d7susypope} agree with (2.1), (2.2) of \cite{Gauntlett:2000ng} (assuming that
\cite{Gauntlett:2000ng} uses $\gamma^{0123456}=+1$ and $\epsilon^{0123456}=+1$), provided \emph{we change the sign of the $S DS$ term} in the Lagrangian (2.1) of
\cite{Gauntlett:2000ng}. It is worth noting, however, that all of the supersymmetric solutions in \cite{Gauntlett:2000ng} had the property $DS=0$ and so this change does not invalidate any of the conclusions of \cite{Gauntlett:2000ng}.}, which we have found to have many typos, using a number of self consistency 
checks, including one we give below.

From these expressions, we can derive the supersymmetry transformations in the $U(1)\times U(1)$ truncation. Following \cite{Liu:1999ai},
we can define
\begin{align}
\hat{\psi}_{\mu}&=\psi_{\mu}-\frac{1}{2}\gamma_{\mu}\Gamma^5\chi_5\,,\quad
\hat{\chi}_{1}=\Gamma^1\chi_1+\frac{3}{2}\Gamma^3\chi_3\,,\quad
\hat{\chi}_{3}=\frac{3}{2}\Gamma^1\chi_1+\Gamma^3\chi_3\,,
\end{align}
and find that the supersymmetry transformations are equivalent\footnote{\label{ftnd7ks}Note that to obtain the Killing spinor equations in
(5.4) of \cite{Ferrero:2021etw}, which had vanishing three-form and $\gamma_{0123456}=-1$, one should set $g_c=1$ and also take $\gamma^\mu\to -\gamma^\mu$.} to
\begin{align}
\begin{split}\label{eq:7d_U(1)2_true_SUSY_variations_1}
\delta\hat{\psi}_{\mu}=&\left[\nabla_{\mu}+\frac{g_c}{2}A^{12}_{\mu}\Gamma^{12}+\frac{g_c}{2}A^{34}_{\mu}\Gamma^{34}-\frac{g_c}{4}e^{-4\lambda_1-4\lambda_2}\gamma_{\mu}+\frac{1}{2}\gamma_{\mu}\gamma^{\nu}\left(\partial_{\nu}\lambda_1+\partial_{\nu}\lambda_2\right)\right.\\
&\left.\phantom{[}-\frac{1}{4}\gamma^\nu\left(e^{-2\lambda_1}\Gamma_{12}F^{12}_{\mu\nu}+e^{-2\lambda_2}\Gamma_{34}F^{34}_{\mu\nu}\right)+\frac{1}{8}\gamma^{\nu\rho}e^{-2\lambda_1-2\lambda_2}\Gamma^5S^5_{\mu\nu\rho}\right]\epsilon\,,
\end{split}
\end{align}
and
\begin{align}
\label{eq:7d_U(1)2_true_SUSY_variations_2}
\delta\hat{\chi}_1=&\left[\frac{1}{4}\left(2\partial_{\mu}\lambda_1+3\partial_{\mu}\lambda_2\right)\gamma^{\mu}+\frac{g_c}{4}\left(e^{2\lambda_2}-e^{-4\lambda_1-4\lambda_2}\right)\right.\nn
&\qquad\qquad\left.\phantom{[}-\frac{1}{16}e^{-2\lambda_2}\Gamma_{34}F^{34}_{\mu\nu}\gamma^{\mu\nu}+\frac{1}{48}\gamma^{\mu\nu\rho}\Gamma^{5}e^{-2\lambda_1-2\lambda_2}S^5_{\mu\nu\rho}\right]\epsilon\,,\nn
\delta\hat{\chi}_3=&\left[\frac{1}{4}\left(3\partial_{\mu}\lambda_1+2\partial_{\mu}\lambda_2\right)\gamma^{\mu}+\frac{g_c}{4}\left(e^{2\lambda_1}-e^{-4\lambda_1-4\lambda_2}\right)\right.\nn
&\qquad\qquad\left.\phantom{[}-\frac{1}{16}e^{-2\lambda_1}\Gamma_{12}F^{12}_{\mu\nu}\gamma^{\mu\nu}+\frac{1}{48}\gamma^{\mu\nu\rho}\Gamma^{5}e^{-2\lambda_1-2\lambda_2}S^5_{\mu\nu\rho}\right]\epsilon\,.
\end{align}
We can provide a highly non-trivial check\footnote{Note that we calculated the commutator in \eqref{pgqeom} assuming 
$\gamma_{0123456}=w$ and $\Gamma_{12345}=v$ where $w,v=\pm 1$. It is only in the case that $w=v=+1$, which
is the conventions we are using, that the commutator gives another supersymmetry transformation when the equations of motion
 are satisfied.}
 of these conditions and the compatibility with the equations of motion
by considering integrability conditions of the supersymmetry transformations as in \cite{MacConamhna:2004fb}.
For example, if we write
$\delta\hat \psi_\mu\equiv \mathcal{D}_\mu\epsilon$ and $\delta\hat \chi_i\equiv \Delta_i\epsilon$ then a long calculation reveals that
\begin{align}\label{intexp}
\gamma^\mu[\mathcal{D}_\mu,\Delta_1]\epsilon&=\Big[-\frac{1}{4}{\ast_7 \mathcal{P}_2}
-\frac{1}{8}e^{2\lambda_2}\left(\ast_7 \mathcal{G}_2\right)_\mu \gamma^\mu \Gamma_{34}
-\frac{w}{48}e^{-2\lambda_1-2\lambda_2}\left(\ast_7 \mathcal{T}\right)_{\mu\nu\rho}\gamma^{\mu\nu\rho} \Gamma^5\Big]\epsilon\nn
&+\Big[-\frac{1}{2}g_c\left(e^{2\lambda_1}+e^{-4\lambda_1-4\lambda_2}\right)+\left(\frac{9}{2}\partial_\mu\lambda_1+5\partial_\mu\lambda_2\right)\gamma^\mu-\frac{3}{8}e^{-2\lambda_1}F^{12}_{\mu\nu}\Gamma_{12}\gamma^{\mu\nu}\nn
&\qquad\qquad\qquad\qquad\qquad\qquad\qquad
+\frac{1}{8}e^{-2\lambda_1-2\lambda_2}S^5_{\mu\nu\rho}\Gamma^5\gamma^{\mu\nu\rho}\Big]\Delta_1\epsilon\nn
&+\Big[\frac{3}{2}g_c\left(-e^{2\lambda_2}+e^{-4\lambda_1-4\lambda_2}\right)+\left(\partial_\mu\lambda_1+\frac{3}{2}\partial_\mu\lambda_2\right)\gamma^\mu-\frac{1}{8}e^{-2\lambda_2}F^{34}_{\mu\nu}\Gamma_{34}\gamma^{\mu\nu}\nn
&\qquad\qquad\qquad\qquad\qquad\qquad\qquad
+\frac{1}{24}e^{-2\lambda_1-2\lambda_2}S^5_{\mu\nu\rho}\Gamma^5\gamma^{\mu\nu\rho}\Big]\Delta_2\epsilon\,,
\end{align}
where $\mathcal{P}_2, \mathcal{G}_2$ and $\mathcal{T}$ were defined in \eqref{pgqeom} and vanish when the equations of motion are satisfied.
Thus, when the equations of motion are satisfied the commutator on the left hand side vanishes for supersymmetric configurations
satisfying $\Delta_i\epsilon=0$.

A final comment is that if one further restricts to a diagonal $U(1)$ sector, by setting $A^{12}=A^{34}$, we can compare with 
the results on minimal $D=7$ gauged supergravity after restricting to a $U(1)\subset SU(2)$ sector of the latter theory. Doing
so we find consistency with e.g. \cite{Passias:2015gya} provided that we set $\gamma_{0123456}=+1$ in \cite{Passias:2015gya}.

\subsection{The fermion reduction}
The explicit $D=7$ Killing spinors for the vacuum $AdS_5\times \spindle_2$ solution were constructed in \cite{Ferrero:2021etw}.
We can use these to determine the $D=5$ supersymmetry variations of bosonic configurations in the consistent truncation
of $D=7$ gauged supergravity on $\spindle_2$. 

Associated with the ansatz for the bosonic fields given in
\eqref{const1}-\eqref{const3}, we introduce the obvious orthonormal frame
\begin{align}
e^{\alpha}=&\,{(yP)^{1/10}}\bar{e}^\alpha\,,\quad
e^5=\frac{y^{3/5}P^{1/10}}{2\sqrt{Q}}dy\,,\quad e^6=\frac{y^{1/10}\sqrt{Q}}{P^{2/5}}\left(d\phi-\frac{4}{3}A_{(1)}\right)\,,
\end{align}
with $\bar{e}^\alpha$, $\alpha=0,1,2,3,4,$ an orthonormal frame for the $D=5$ metric $ds^2_5$.
It is convenient to use an explicit set of  $D = 7$ gamma matrices $\gamma^\mu$, associated with the decomposition $SO(1,6)\to SO(1,4)\times SO(2)$,  given by
\begin{align}\label{7dgam5dgam}
\gamma_{\alpha}=\beta_{\alpha}\otimes\sigma^3\,,\quad \gamma_{5}=\mathds{1}\otimes\sigma^1\,,\quad \gamma_{6}=\mathds{1}\otimes\sigma^2\,,
\end{align}
where $\beta_\alpha$ are $D=5$ gamma matrices satisfying $\beta_{01234}=-i\mathds{1}$. For these $D=5$ gamma matrices, we can define $B_5$ satisfying $B_5\beta_\alpha B_5^{-1}=-\beta_\alpha^*$ and $B_5^2=1$. We can also define $B_2=\sigma^1$, so that
$B_2(\sigma^1,\sigma^2) B_2^{-1}=+(\sigma^1,\sigma^2)^*$. We then define $B_7=B_5\otimes B_2$ with 
$B_7\gamma^\mu B_7^{-1}=+\gamma^{\mu *}$.

We now consider a $D=7$ spinor $e^{-\frac{3i\phi}{4}}\varepsilon\otimes\zeta_{(2)}$, where $\varepsilon$ is an arbitrary $D=5$ spinor and
the two component spinor $\zeta_{(2)}$ on the spindle $\spindle_2$ is given below. We will also need
the $D=7$ conjugate spinor which is given by $e^{+\frac{3i\phi}{4}}\varepsilon^c\otimes\zeta^c_{(2)}$, where $\varepsilon^c\equiv B_5\varepsilon^*$ and $\zeta_{(2)}^c\equiv \sigma^1\zeta_{(2)}^*$. We then 
consider the following ansatz for $D=7$ Killing spinors of $SO(5)$ gauged supergravity
\begin{align}\label{ksanstazconskk}
\epsilon=e^{-\frac{3i\phi}{4}}\varepsilon\otimes\zeta_{(2)}\otimes u_-\qquad \text{or}\qquad \epsilon=e^{\frac{3i\phi}{4}}\varepsilon^c\otimes\zeta_{(2)}^c\otimes u_+\,,
\end{align}
where $u_\pm$ are two four-component spinors acted on by the $SO(5)$ gamma matrices $\Gamma^i$ which have the same eigenvalue with respect
to both $\Gamma^{12}$ and $\Gamma^{34}$:
\begin{align}
\Gamma^{12}u_\pm=\Gamma^{34}u_\pm=\pm iu_\pm\,.
\end{align}
The two-component spinor $\zeta_{(2)}$ on the spindle $\spindle_2$ is given by\footnote{Note that these are not exactly the same as those given
in \cite{Ferrero:2021etw} for the $AdS_5\times \spindle_2$ solution, due to the different supersymmetry conventions as noted in footnote \ref{ftnd7ks}.}
\begin{align}
\zeta_{(2)}=\frac{y^{1/20}}{\sqrt{2}P^{1/5}}\left(
\begin{array}{c}
 \sqrt{\sqrt{P}+2y^{3/2}} \\
 \sqrt{\sqrt{P}-2y^{3/2}}    \\
\end{array}
\right)\,,\quad 
\zeta^c_{(2)}=\frac{y^{1/20}}{\sqrt{2}P^{1/5}}\left(
\begin{array}{c}
 \sqrt{\sqrt{P}-2y^{3/2}} \\
 \sqrt{\sqrt{P}+2y^{3/2}}    \\
\end{array}
\right)\,, 
\end{align}
satisfying
\begin{align}
\label{eq:nice_way_to_write_projection}
\left(\cos\alpha\gamma^{56}+ i\sin\alpha\gamma^5\right)\varepsilon\otimes\zeta_{(2)}=&\,+i\varepsilon\otimes\zeta_{(2)}\,,\nn
\Leftrightarrow\quad \left(\cos\alpha\gamma^{56}- i\sin\alpha\gamma^5\right)\varepsilon^c\otimes\zeta_{(2)}^c=&\,-i\varepsilon^c\otimes\zeta_{(2)}^c\,,
\end{align}
where
\begin{align}
\begin{split}
\cos\alpha=\frac{2y^{3/2}}{\sqrt{P}}\,,\qquad \sin\alpha=2\sqrt{\frac{Q}{P}}\,.
\end{split}
\end{align}
The explicit phase appearing in \eqref{ksanstazconskk} arises because of the specific gauge that we are using for the gauge fields.

We next substitute this ansatz into the $D=7$ Killing spinor equations 
\eqref{eq:7d_U(1)2_true_SUSY_variations_1}-\eqref{eq:7d_U(1)2_true_SUSY_variations_2}. We find that $\delta\hat\chi_1=\delta\hat\chi_3=0$. To analyse 
$\delta\hat{\psi}_{\mu}$ we need the spin connection associated with the frame which has the form
\begin{align}
\begin{split}
\omega^{\alpha}_{\phantom{\alpha}\beta}=&\,\bar{\omega}^{\alpha}_{\phantom{\alpha}\beta}+\frac{2\sqrt{Q}}{3y^{1/10}P^{3/5}}F^{\alpha}_{\phantom{\alpha}\beta}\,e^6\,,\quad \omega^{\alpha}_{\phantom{\alpha}5}=\frac{2\sqrt{Q}}{y^{7/10}P^{1/5}}\partial_y\left[(yP)^{1/10}\right]e^\alpha\,,\\
\omega^{6}_{\phantom{6}\alpha}=&-\frac{2\sqrt{Q}}{3y^{1/10}P^{3/5}}F_{\alpha\beta}\,e^\beta\,,\qquad \qquad\omega^{6}_{\phantom{6}5}=\frac{2P^{3/10}}{y^{7/10}}\partial_y\left(\frac{y^{1/10}\sqrt{Q}}{P^{2/5}}\right)e^{6}\,,
\end{split}
\end{align}
where $\bar{\omega}^{\alpha}_{\phantom{\alpha}\beta}$ is the spin connection for the $D=5$ metric $ds^2_5$.
After some work, we find that $\delta\hat{\psi}_{\mu}=0$ is equivalent, in either of the two cases in \eqref{ksanstazconskk}, to the $D=5$ spinor $\varepsilon$ satisfying
\begin{align}
\begin{split}\label{eq:gravitino_variation_5d_minimal} 
\left[\bar{\nabla}_{\alpha}- \frac{1}{2}\beta_{\alpha}- i A_{\alpha}-\frac{i}{12}\left(\beta_{\alpha}^{\phantom{\alpha}\beta\rho}-4\delta_{\alpha}^\beta\beta^{\rho}\right)F_{\beta\rho}\right]\varepsilon=0\,,
\end{split}
\end{align}
and recall we have $\beta_{01234}=-i\mathds{1}$.
This is precisely the Killing spinor equation for a bosonic configuration of $D=5$ gauged supergravity satisfying
the equations of motion \eqref{D5EOM} with $g_c=1$.

\subsection{R-symmetry of $AdS_3\times\spindle_1\ltimes\spindle_2$ solution}
Here we identify the R-symmetry of the $AdS_3\times\spindle_1\ltimes\spindle_2$ solution of $D=7$ gauged supergravity by
constructing a suitable Killing spinor bi-linear.

For the decomposition $SO(1,4)\to SO(1,2)\times SO(2)$ we write the
$D=5$ gamma matrices as
$\beta_a=\alpha_a\otimes\sigma^3$, $\beta_3=\mathds{1}\otimes\sigma^2$ and $\beta_4=\mathds{1}\otimes\sigma^1$, where $\alpha_a$ are $D=3$ gamma matrices 
satisfying $\alpha_0\alpha_1\alpha_2=\mathds{1}$, for example 
$\alpha_0=i\sigma^2$,
$\alpha_1=\sigma^1$ and $\alpha_2=\sigma^3$. In this basis, we can take $B_5=\beta_3$.

The $D=5$ Killing spinors solving \eqref{eq:gravitino_variation_5d_minimal} for the $AdS_3\times \spindle_1$ solution of \cite{Ferrero:2020laf}, given in \eqref{soln}-\eqref{Sigmametric},
can be written as
\begin{equation}\label{deefiveks}
\varepsilon=\vartheta_{AdS_3}\otimes \zeta_{(1)}\,,
\end{equation}
where $\vartheta_{AdS_3}$ is a Killing spinor on $AdS_3$ satisfying $\nabla_a\vartheta_{AdS_3}=\frac{1}{2}\alpha_a\vartheta_{AdS_3}$,
and $\zeta_{(1)}$ is a spinor on the spindle $\spindle_1$ given by
\begin{equation}\label{eq:spindle_spinor}
\zeta_{(1)}=\left(\frac{\sqrt{f_1(x)}}{\sqrt{x}},i\frac{\sqrt{f_2(x)}}{\sqrt{x}}\right)\,,
\end{equation}
where
\begin{equation}
f_1(x)=-a+2x^{3/2}+3x\,,\quad f_2(x)=a+2x^{3/2}-3x\,,
\end{equation}
which satisfy $f(x)=f_1(x)f_2(x)$, with $f(x)$ given in \eqref{efffn}. As in \cite{Ferrero:2020laf}, the spinor is independent of the coordinate $\psi$ on the spindle associated with the specific gauge used in \eqref{soln}.

 We now provide the explicit expression of the Killing spinors on $AdS_3$. We 
write the metric on $AdS_3$ in Poincar\'e coordinates as
\begin{align}
ds^2(AdS_3)=\frac{-(dx^0)^2+(dx^1)^2+dr^2}{r^2}\, ,
\end{align}
and then from e.g. appendix B of \cite{Ferrero:2021etw}, we can write 
\begin{align}
\begin{split}
\vartheta_{AdS_3}^{(1)}&=\frac{1}{\sqrt{r}}\left(\begin{array}{c}
0\\
1 \\
\end{array}\right)\,,\quad  \vartheta_{AdS_3}^{(2)}=\frac{1}{\sqrt{r}}\left[ix^0\sigma^2+x^1\sigma^1+r\sigma^3\right]\left(\begin{array}{c}
1 \\
0 \\
\end{array}\right)\,,
\end{split}
\end{align}
associated with the Poincar\'e and superconformal Killing spinors, respectively.
Overall, the $D=7$ Killing spinors in \eqref{ksanstazconskk} are given by
\begin{align}\label{ksanstazconskkagain}
\epsilon^{(i)}=e^{-\frac{3i\phi}{4}}\vartheta_{AdS_3}^{(i)}\otimes\zeta_{(1)}\otimes\zeta_{(2)}\otimes u_-\quad \text{or}\quad \epsilon^{(i)}=e^{\frac{3i\phi}{4}}\vartheta_{AdS_3}^{(i)}\otimes\sigma^1\zeta_{(1)}^*\otimes\zeta_{(2)}^c\otimes u_+\,.
\end{align}
Notice in particular that there are two Poincar\'e Killing spinors of the same chirality associated with a $d=2$, $\mathcal{N}=(0,2)$ 
SCFT. We can now consider bi-linears to extract the associated superconformal algebra. We have $u_-^\dagger u_+=0$ and
we normalise $u_-^\dagger u_-=u_+^\dagger u_+=6$ for convenience. Defining $\bar{\epsilon}=\epsilon^\dagger \gamma_0$
for bi-linears of the $u_-$ Killing spinors we obtain
\begin{align}
&\left[\bar{\epsilon}^{(1)}\gamma^m\epsilon^{(1)}\right] \partial_m=\partial_{x^0}-\partial_{x^1}\,,\nn
&\left[\bar{\epsilon}^{(2)}\gamma^m\epsilon^{(2)}\right] \partial_m=\left[r^2+(x^0-x^1)^2\right]\partial_{x^0}+\left[r^2-(x^0-x^1)^2\right]\partial_{x^1}+2r(x^0-x^1)\partial_{r}\,,\nn
&\left[\bar{\epsilon}^{(1)}\gamma^m\epsilon^{(2)}\right] \partial_m=-(x^0\partial_{x^0}+x^1\partial_{x^1}
+r\partial_{r})
+(x^1\partial_{x^0}+x^0\partial_{x^1})-i\Big(2\partial_{\psi}-\frac{4}{3}\partial_{\phi}\Big)\,,
\end{align}
and we obtain the same result for bi-linears of the $u_+$ Killing spinors.
On the right hand side of this expression we have precisely the Killing vectors generating the $d=2$, $\mathcal{N}=(0,2)$ superconformal algebra with
the R-symmetry Killing vector identified as
\begin{align}\label{cankvbilin}
R=2\partial_{\psi}-\frac{4}{3}\partial_{\phi}\,.
\end{align}

\section{Anomaly polynomial for $\spindle_1\ltimes\spindle_2$}\label{app:c}

We are interested in determining the anomaly polynomial associated with $N$ M5-branes wrapped on 
$\mathbb{R}^{1,1}\times\spindle_1\ltimes\spindle_2$. The $D=7$ supergravity construction shows that
we are interested in activating background gauge fields in a $U(1)\times U(1)\subset SO(5)_R$ subgroup of the $SO(5)_R$
symmetry of the M5-brane worldvolume theory. This setup is associated with the normal bundle $\mathcal{N}$ to the M5-branes splitting
via $\mathcal{N}=\mathbb{R}\oplus\mathcal{N}_1\oplus\mathcal{N}_2$, where $\mathcal{N}_i$ are complex line bundles. In the large $N$ limit, we can then write the anomaly polynomial as the eight-form (e.g. \cite{Hosseini:2020vgl}):
\begin{align}\label{6danompoly}
\mathcal{A}_{6d}&=\frac{N^3}{24}c_1(\mathcal{N}_1)^2c_1(\mathcal{N}_2)^2\,.
\end{align}

In compactifying the $d=6$ theory on $\spindle_1\ltimes\spindle_2$, we need to take into account the $U(1)_{J_1}\times U(1)_{J_2}$
global symmetry that arises from the isometries of $\spindle_1\ltimes\spindle_2$. Generalising \cite{Ferrero:2020laf,Ferrero:2021wvk,Hosseini:2020vgl} we then want to compute the 
$d=6$ anomaly polynomial \eqref{6danompoly} on an eight-manifold, $Z_8$, which is defined as the total space of a $\spindle_1\ltimes\spindle_2$ 
fibration over a four-manifold: 
\begin{align}
\spindle_1\ltimes\spindle_2 \hookrightarrow Z_8 \hookrightarrow Z_4\,.
\end{align}
As in \cite{Ferrero:2020laf,Ferrero:2021wvk} we demand that the Killing spinor is invariant under the 
$U(1)_{J_1}\times U(1)_{J_2}$ symmetry generated by the normalised Killing vectors $(\frac{\Delta \psi}{2\pi}\partial_\psi,\frac{\Delta \phi}{2\pi}\partial_\phi)$. 

Now recall the gauge fields of the $D=7$ supergravity solution given in \eqref{d7sssol2}.
In this gauge, from \eqref{ksanstazconskk} and \eqref{deefiveks}, we see that the Killing spinors depend on $\phi$ but are independent
of $\psi$. We want to work in a gauge in which there is no dependence on either $\phi$ or $\psi$. We therefore
consider gauge fields of the form
\begin{align}
A^{12}_{(1)}&=\left(\frac{q_1}{h_1}-1+\mathfrak{a}_1\right)d\phi+\left[\mathfrak{b}_1-\frac{1}{3}\left(\frac{q_1}{h_1}-1\right)\left(1-\frac{a}{x}\right)\right]d\psi\,,\nn
A^{34}_{(1)}&=\left(\frac{q_2}{h_2}-1+\mathfrak{a}_2\right)d\phi+\left[\mathfrak{b}_2-\frac{1}{3}\left(\frac{q_2}{h_2}-1\right)\left(1-\frac{a}{x}\right)\right]d\psi\,,
\end{align}
where we have allowed gauge transformations parametrised by
$\mathfrak{a}_i, \mathfrak{b}_i$ with $i=1,2$ satisfying
$\mathfrak{a}_1+\mathfrak{a}_2=\frac{3}{2}$ and $\mathfrak{b}_1+\mathfrak{b}_2=0$.
Following the procedure in \cite{Ferrero:2021wvk,Ferrero:2020laf}, we now introduce the following connection one-forms on $Z_8$:
\begin{align}
\begin{split}
\mathcal{A}^1_{(1)}&=\rho_1(y)\left(d\phi+\frac{\Delta \phi}{2\pi}A_{J_2}\right)+\left[\mathfrak{b}_1-\frac{1}{3}\theta_1(y)\left(1-\frac{a}{x}\right)\right]\left(d\psi+\frac{\Delta \psi}{2\pi}A_{J_1}\right)\,,\\
\mathcal{A}^2_{(1)}&=\rho_2(y)\left(d\phi+\frac{\Delta \phi}{2\pi}A_{J_2}\right)+\left[\mathfrak{b}_2-\frac{1}{3}\theta_2(y)\left(1-\frac{a}{x}\right)\right]\left(d\psi+\frac{\Delta \psi}{2\pi}A_{J_1}\right)\,,
\end{split}
\end{align}
where we have defined two functions on the fibre $\spindle_2$ of $\spindle_1\ltimes\spindle_2$:
\begin{align}
\theta_i(y)&=\frac{q_i}{h_i(y)}-1\,,\nn
\rho_i(y)&=\theta_i(y)+\mathfrak{a}_i\,,
\end{align}
with $\rho_i'=\theta_i'$ and $(A_{J_1},A_{J_2})$ are connection one-forms associated with the $U(1)_{J_1}\times U(1)_{J_2}$ symmetry.
The associated curvature two-forms $\mathcal{F}^i_{(2)}=d\mathcal{A}^i_{(1)}$
are given by
\begin{align}
\mathcal{F}^i_{(2)}&=\rho'_idy\wedge\left[d\phi+\frac{\Delta \phi}{2\pi}A_{J_2}-\frac{1}{3}\left(1-\frac{a}{x}\right)\left(d\psi+\frac{\Delta \psi}{2\pi}A_{J_1}\right)\right]+\frac{\rho_i\Delta \phi}{2\pi}F_{J_2}\nn
&\quad-\frac{a\theta_i}{3x^2}dx\wedge \left(d\psi+\frac{\Delta \psi}{2\pi}A_{J_1}\right)+\frac{\Delta \psi}{2\pi}\left[\mathfrak{b}_i-\frac{1}{3}\theta_i\left(1-\frac{a}{x}\right)\right]F_{J_1}\,,
\end{align}
with $F_{J_i}=d A_{J_i}$. We have $[\mathcal{F}^i_{(2)}/2\pi]\in H^2(Z_8,\mathbb{Z})$  and we have normalised so that $c_1(J_i)\equiv [F_{J_i}/2\pi]\in H^2(Z_4,\mathbb{Z})$.

We now write 
\begin{align}
c_1(\mathcal{N}_i)=\Delta_i c_1(R_{2d})+c_1(\mathcal{F}^i_{(2)})\,,
\end{align}
where $R_{2d}$ is the pull-back of a $U(1)_R$ symmetry bundle over $Z_4$ and the 
trial R-charges satisfy $\Delta_1+\Delta_2=2$. This latter condition ensures that the preserved spinor has R-charge 1.
The $d=2$ anomaly polynomial on $Z_4$, at large $N$, is now obtained by 
substituting $c_1(\mathcal{N}_i)$ into $\mathcal{A}_{6d}$ given in \eqref{6danompoly} and then integrating over $\spindle_1\ltimes\spindle_2$,
\begin{align}
\mathcal{A}_{2d}&=\frac{N^3}{24}\int_{\spindle_1\ltimes\spindle_2}c_1(\mathcal{N}_1)^2c_1(\mathcal{N}_2)^2\,.
\end{align}
This gives the four-form
\begin{align}\label{a2dpol}
\mathcal{A}_{2d}&=\frac{N^3}{24}\Big\{ 
(\Delta_1^2 I_1+\Delta_2^2I_2+\Delta_1\Delta_2 I_3)c_1(R_{2d})^2 
+(\Delta_1 I_4+\Delta_2I_5)c_1(R_{2d})c_1({J_1})\nn
&+(\Delta_1 I_6+\Delta_2I_7)c_1(R_{2d})c_1({J_2})
+I_{8}c_1({J_1})^2+I_{9}c_1({J_2})^2+I_{10}c_1({J_1})c_1({J_2})\Big\}\,,
\end{align}
where
\begin{align}
I_{1}&= \frac{\Delta \phi}{3\pi}\left(\frac{1}{m_-}-\frac{1}{m_+}\right)\left[\theta_2^2\right]^{y_{3}}_{y_{2}}  \,, \nn
I_{2}&=  \frac{\Delta \phi}{3\pi}\left(\frac{1}{m_-}-\frac{1}{m_+}\right)\left[\theta_1^2\right]^{\y_{3}}_{\y_{2}}  \,, \nn
I_{3}&= \frac{\Delta \phi}{3\pi}\left(\frac{1}{m_-}-\frac{1}{m_+}\right) \left[4\theta_1\theta_2\right]^{\y_{3}}_{\y_{2}} \,,    \nn
I_{4}&= \frac{4\Delta \phi}{9\pi}\left(\frac{1}{m_-^2}-\frac{1}{m_+^2}\right)\left[\theta_1\theta_2^2\right]^{\y_{3}}_{\y_{2}}
+ \frac{8\Delta \phi}{9\pi}\frac{m_+^3-m_-^3}{m_-^2m_+^2(m_-+m_+)}\left(\frac{\mathfrak{b}_1}{2}\left[\theta_2^2\right]^{\y_{3}}_{\y_{2}}+{\mathfrak{b}_2}\left[\theta_1\theta_2\right]^{\y_{3}}_{\y_{2}}\right) \,,\nn
I_{5}&= \frac{4\Delta \phi}{9\pi}\left(\frac{1}{m_-^2}-\frac{1}{m_+^2}\right)\left[\theta_1^2\theta_2\right]^{\y_{3}}_{\y_{2}} 
+  \frac{8\Delta \phi}{9\pi}\frac{m_+^3-m_-^3}{m_-^2m_+^2(m_-+m_+)}\left(
{\mathfrak{b}_1}\left[\theta_1\theta_2\right]^{\y_{3}}_{\y_{2}}
+\frac{\mathfrak{b}_2}{2}\left[\theta_1^2\right]^{\y_{3}}_{\y_{2}}\right)\,,
\end{align}
and
\begin{align}
I_{6}&=\frac{2(\Delta \phi)^2}{3\pi^2}\left(\frac{1}{m_-}-\frac{1}{m_+}\right)\left(\left[\rho_1\rho_2\theta_2\right]^{\y_{3}}_{\y_{2}}-\frac{\mathfrak{a}_1}{2}\left[\rho_2^2\right]^{\y_{3}}_{\y_{2}}\right) \,,  \nn
I_{7}&= \frac{2(\Delta \phi)^2}{3\pi^2}\left(\frac{1}{m_-}-\frac{1}{m_+}\right)\left(\left[\rho_1\rho_2\theta_1\right]^{\y_{3}}_{\y_{2}}-\frac{\mathfrak{a}_2}{2}\left[\rho_1^2\right]^{\y_{3}}_{\y_{2}}\right) \,,\nn
I_{8}&= \frac{4\Delta \phi}{27\pi}\left(\frac{1}{m_-^3}-\frac{1}{m_+^3}\right)\left(\left[\theta_1^2\theta_2^2\right]^{\y_{3}}_{\y_{2}}+2\mathfrak{b}_1\left[\theta_1\theta_2^2\right]^{\y_{3}}_{\y_{2}}+2\mathfrak{b}_2\left[\theta_1^2\theta_2\right]^{\y_{3}}_{\y_{2}}\right)  \nn
&+ \frac{8\Delta \phi}{27\pi}\frac{(m_-^2+m_-m_++m_+^2)(m_+^3-m_-^3)}{m_-^3m_+^3(m_-+m_+)^2}\left(\frac{\mathfrak{b}_1^2}{2}\left[\theta_2^2\right]^{y_{2,3}}_{y_{2,2}}
+\frac{\mathfrak{b}_2^2}{2}\left[\theta_1^2\right]^{y_{2,3}}_{y_{2,2}}
+2\mathfrak{b}_1\mathfrak{b}_2\left[\theta_1\theta_2\right]^{y_{2,3}}_{y_{2,2}}\right)\,,\nn
I_{9}&=  \frac{(\Delta \phi)^3}{6\pi^3}\left(\frac{1}{m_-}-\frac{1}{m_+}\right)\left(\frac{3}{2}\left[\rho_1^2\rho_2^2\right]^{\y_{3}}_{\y_{2}}-{\mathfrak{a}_1}\left[\rho_1\rho_2^2\right]^{\y_{3}}_{\y_{2}}-{\mathfrak{a}_2}\left[\rho_1^2\rho_2\right]^{\y_{3}}_{\y_{2}}\right)\,, \nn
I_{10}&= \frac{2(\Delta \phi)^2}{9\pi^2}\left(\frac{1}{m_-^2}-\frac{1}{m_+^2}\right)\left(\frac{3}{2}\left[\theta_1^2\theta_2^2\right]^{\y_{3}}_{\y_{2}}+{\mathfrak{a}_1}\left[\theta_1\theta_2^2\right]^{\y_{3}}_{\y_{2}}+{\mathfrak{a}_2}\left[\theta_1^2\theta_2\right]^{\y_{3}}_{\y_{2}}\right)  \nn
&+\frac{4(\Delta \phi)^2}{9\pi^2}\frac{(m_+^3-m_-^3)}{m_-^2m_+^2(m_-+m_+)}\left(\mathfrak{b}_1\left[\left[\rho_1\rho_2\theta_2\right]^{\y_{3}}_{\y_{2}}-\frac{\mathfrak{a}_1}{2}\left[\rho_2^2\right]^{\y_{3}}_{\y_{2}}\right]+\mathfrak{b}_2\left[\left[\rho_1\rho_2\theta_1\right]^{\y_{3}}_{\y_{2}}-\frac{\mathfrak{a}_2}{2}\left[\rho_1^2\right]^{\y_{3}}_{\y_{2}}\right]\right) \,.
\end{align}

Having obtained the $d=2$ anomaly polynomial, we can derive the associated $d=2$ central charge using
$c$-extremization \cite{Benini:2012cz}. The coefficient of $\tfrac{1}{2}c_1(L_a)c_1(L_b)$ in the expression for
$\mathcal{A}_{2d}$ given in \eqref{a2dpol} is $\text{tr}\gamma^3Q_a Q_b$
where the global symmetry $Q_a$ is associated with the $U(1)$ bundle $L_a$ over $Z_4$ and $\gamma^3$ is the $d=2$ chirality operator. Now $c$-extremization implies that the $d=2$ superconformal $U(1)_R$ symmetry extremises
\begin{align}
c_{trial}=3 \text{tr}\gamma^3 R_{trial}^2\,,
\end{align}
over the space of possible R-symmetries. We write the trial R-symmetry as
\begin{align}
R_{trial}=R_{2d}+\epsilon_1J_1+\epsilon_2 J_2\,,
\end{align}
which leads to a trial central charge given by 
\begin{align}
c_{trial}=-{6}\frac{N^3}{24}\Big\{ 
&(\Delta_1^2 I_1+\Delta_2^2I_2+\Delta_1\Delta_2 I_3) 
+(\Delta_1 I_4+\Delta_2I_5)\epsilon_1
+(\Delta_1 I_6+\Delta_2I_7)\epsilon_2\nn
&+I_{8}\epsilon_1^2+I_{9}\epsilon_2^2+I_{10}\epsilon_1\epsilon_2\Big\}\,.
\end{align}
The trial R-symmetry is parametrised by $\epsilon_1,\epsilon_2$ and $\Delta_1,\Delta_2$ subject to $\Delta_1+\Delta_2=2$ as above.
We also have dependence on the gauge parameters $\mathfrak{a}_i, \mathfrak{b}_i$ which 
we recall satisfy $\mathfrak{a}_1+\mathfrak{a}_2=\frac{3}{2}$ and $\mathfrak{b}_1+\mathfrak{b}_2=0$ .

Carrying out the extremization, we find that at the critical point
\begin{align}
\begin{split}
\epsilon^*_1&=\frac{3m_-m_+(m_-+m_+)}{m_-^2+m_-m_++m_+^2}=2\frac{2\pi}{\Delta\psi}\,,\nn
\epsilon^*_2&=-\frac{n_-n_+(2n_--p_1-p_2)(\mathtt{s}+p_1+p_2)}{(n_--p_1)(p_2-n_-)(\mathtt{s}+2p_1+2p_2)}=-\frac{4}{3}\frac{2\pi}{\Delta\phi}\,,
\end{split}
\end{align}
and
\begin{align}
\begin{split}
\Delta_1^*=\frac{4\mathfrak{a}_1}{3}-2\mathfrak{b}_1\,,\quad\Delta_2^*=\frac{4\mathfrak{a}_2}{3}-2\mathfrak{b}_2=2-\Delta_1^*\,,
\end{split}
\end{align}
with the corresponding central charge given by
\begin{align}
\begin{split}
c^*
&=\frac{N^3}{2}\frac{(m_--m_+)^3}{m_-m_+\left(m_-^2+m_-m_++m_+^2\right)}\frac{p_1^2p_2^2(\mathtt{s}+p_1+p_2)}{n_-n_+(n_--p_1)(p_2-n_-)(\mathtt{s}+2p_1+2p_2)^2}\,.
\end{split}
\end{align}
This expression for the central charge is in exact agreement with the supergravity result \eqref{finalcc}.
We can also compare the twisting of the R-symmetry that arises from the two global $U(1)$ symmetries, $J_i$.
We can identify $J_1, J_2$ with $\partial_{\tilde\psi}$, $\partial_{\tilde\phi}$, respectively, where $\tilde\psi\equiv (2\pi/\Delta\psi)\psi$,
 $\tilde\phi\equiv (2\pi/\Delta\phi)\phi$ with $\Delta\tilde\psi=\Delta\tilde\phi=2\pi$. Then at the extremal point we have
 \begin{align}
 \epsilon_1^*J_1+\epsilon_2^* J_2= 2\partial_\psi-\frac{4}{3}\partial_\phi\,,
 \end{align}
and this is precisely the Killing vector that appears as a Killing spinor bilinear in the supergravity solution \eqref{cankvbilin}.

It is interesting to highlight that $\epsilon^*_1,\epsilon^*_2$ and $c^*$ are all independent of the gauge parameters  
$\mathfrak{a}_i, \mathfrak{b}_i$. We also observe that there is a one-parameter family of preferred
gauges with $(4/3)\mathfrak{a}_1-2\mathfrak{b}_1=1$ for which
\begin{align}
\Delta_1^*=\Delta_2^*=1\,.
\end{align}
This includes as a special case a symmetric gauge with $\mathfrak{a}_1=\mathfrak{a}_2=\frac{3}{4}$ 
and $\mathfrak{b}_1=\mathfrak{b}_2=0$.


\providecommand{\href}[2]{#2}\begingroup\raggedright\endgroup

\end{document}